\documentclass[aip,amsmath,twocolumn,10pt,jcp]{revtex4-1}
\usepackage{bm}%
\usepackage[colorlinks=true,linkcolor=blue]{hyperref}%
\usepackage{graphicx}
\usepackage{dcolumn}
\usepackage{subfigure}
\usepackage{amsmath}
\usepackage{amsfonts}
\usepackage{ragged2e}
\usepackage{amssymb}
\usepackage{geometry}
\usepackage{graphicx}
\usepackage{xcolor}

\expandafter\ifx\csname package@font\endcsname\relax\else
 \expandafter\expandafter
 \expandafter\usepackage
 \expandafter\expandafter
 \expandafter{\csname package@font\endcsname}%
\fi
\begin{document}
	
	\title{Building machine learning force fields for nanoclusters}

	\author{Claudio Zeni} 
	\email{claudio.zeni@kcl.ac.uk}
	\affiliation{Department of Physics, King's College London, Strand, London WC2R 2LS, United Kingdom}
	
	\author{Kevin Rossi} 
	\email{kevin.rossi@kcl.ac.uk}
	\affiliation{Department of Physics, King's College London, Strand, London WC2R 2LS, United Kingdom}
	
	\author{Aldo Glielmo}
	\affiliation{Department of Physics, King's College London, Strand, London WC2R 2LS, United Kingdom}
	
	\author{\'Ad\'am Fekete}
	\affiliation{Department of Physics, King's College London, Strand, London WC2R 2LS, United Kingdom}
	
	\author{Nicola Gaston}
	\affiliation{MacDiarmid Institute for Advanced Materials and Nanotechnology; University of Auckland, Private Bag 92019, Auckland 1010, New Zealand}
	
	\author{Francesca Baletto}
	\affiliation{Department of Physics, King's College London, Strand, London WC2R 2LS, United Kingdom}
	
	\author{Alessandro De Vita}
	\affiliation{Department of Physics, King's College London, Strand, London WC2R 2LS, United Kingdom} 
	\affiliation{Dipartimento di Ingegneria e Architettura, Universit\`{a} di Trieste, via A. Valerio 2, I-34127 Trieste, Italy}

	\date{February 2018}%
	\revised{June 2018}%
	\begin{abstract}
	We assess Gaussian process (GP) regression as a technique to model interatomic forces in metal nanoclusters by analysing the performance of 2-body, 3-body and many-body kernel functions on a set of 19-atom Ni cluster structures.  
	We find that 2-body GP kernels fail to provide faithful force estimates, despite succeeding in bulk Ni systems.
    However, both 3- and many-body kernels predict forces within a $\sim$0.1 eV/$\text{\AA}$ average error even for small training datasets, and achieve high accuracy even on out-of-sample, high temperature, structures.
	While training and testing on the same structure always provides satisfactory accuracy,  cross-testing on dissimilar structures leads to higher prediction errors, posing an extrapolation problem.   
	This can be cured using heterogeneous training on databases that contain more than one structure, which results in a good trade-off between versatility and overall accuracy. 
	Starting from a 3-body kernel trained this way, we build an efficient non-parametric 3-body force field that allows accurate prediction of structural properties at finite temperatures, following a newly developed scheme [Glielmo et al. PRB 97, 184307 (2018)].  
	We use this to assess the thermal stability of Ni$_{19}$ nanoclusters at a fractional cost of full \emph{ab initio} calculations.
	\end{abstract}
	\maketitle
	
	\section{Introduction}
	
	Metallic nanoparticles have fascinating chemo-physical properties, different from those of their individual atomic constituents and their bulk counterparts\cite{tyo2015catalysis, li2012investigation, billas1994magnetism, di2016geometrical, cottancin2006optical}.
	Because of the variety of isomers accessible at finite temperatures and the lack of translational symmetry, implying a non-trivial interplay between their geometric and electronic structure, a comprehensive understanding of metallic nanoclusters remains challenging, despite their potential use in many advanced applications \cite{steenbergen2012electronic, ojha2013single, steenbergen2013first, steenbergen2015quantum, ojha201520, steenbergen2015two, vargas2009fluxionality, pavan2013sampling, santarossa2010free, zhai2016ensemble, zhai2017fluxionality}.
	Density Functional Theory (DFT) is the most common framework to investigate the static and dynamical properties of nanoclusters of few tens of atoms, for which standard classical force fields cannot systematically be relied upon to provide sufficient accuracy.
	However, DFT-based  calculations are very expensive, and probing limited timescales in first principles dynamical simulations may lead to poor sampling of the nanoclusters' configuration space.
	Machine Learning Force Fields (ML-FFs) may provide a solution to this problem, and the needed access to the dynamical properties of nanoclusters, by extending by several orders of magnitude the accessible time scale, while still describing sufficiently accurately the interactions between the cluster atoms.
	The ML-FFs of most widespread use in cluster science are based on artificial neural networks \cite{Behler:2007fe}.
	In many works aiming at investigating nanoclusters' properties the training databases were constructed from clusters of several sizes, involving structures based on different Bravais lattices and surfaces with different crystallographic orientations \cite{artrith2013neural, artrith2014understanding, artrith2015grand, ouyang2015global, chiriki2016modeling, chiriki2017neural} .  
	The \emph{ex novo} production of such databases requires many expensive quantum calculations: while some redundancy is hard to avoid, the neural network architecture, by means of multiple layers and a high number of fitted parameters, is usually able to extract the necessary information and correctly predict energies and forces in specific scenarios.
	The resulting trained force-field, although versatile, can be however difficult to interpret because of the inherent complexity of the algorithm.	

	Another commonly used class of ML-FFs is based on Gaussian Process (GP) regression \cite{Bartok:2010fj, Glielmo:2017dj}, and has recently been used to predict properties of both bulk \cite{suzuki2017machine, Deringer:2017ea} and molecular \cite{uteva2017interpolation, cui2016efficient, miwa2016molecular} systems.
	While GPs have been also applied to predict adsorption energies of small molecules on NiGa and RhAu nanoclusters \cite{ulissi2017machine, jinnouchi2017predicting}, they have never been used so far to estimate the finite-temperature structural properties of a nanoparticle.
	GPs are usually easier to interpret as they contain a small number of physically meaningful hyperparameters.
Moreover, including symmetries such as the translational invariance and rotational covariance of forces\cite{Glielmo:2017dj} or choosing simplifying approximations such as selecting the $n$-body order of interaction between atoms\cite{Glielmo2018, bartok2013machine, shapeev2016moment, Deringer:2017ea} in the algorithm is straightforward in the case of GPs, where these properties can be encoded the kernel function, enabling fast training and high prediction accuracy.

	In this work we systematically asses the GP force estimates for a set of Ni$_{19}$ nanocluster structures, as obtained from  2-, 3-, and many-body kernels, and for a number of training databases of different nature and size. 
	Consistent with the significant change of physical properties occurring during the atom-to-bulk transition, we find that a 2-body kernel generally fails to correctly predict forces in small-sized Ni nanoclusters, despite doing so for bulk Ni systems. 
	However, a 3-body kernel performs well, hinting to an increased significance of angular terms in the bonding.\\

	 The choice of training dataset is key to the performance of any ML-FF. 
		The search for an optimal training dataset which may encompass all the relevant structures while avoiding redundancy is therefore of interest, to guarantee accuracy while limiting the need to produce \emph{ad-hoc} \emph{ab-initio} databases.
	Consistent with intuition, the GP accuracy decreases as the structural dissimilarity increases between the training and testing morphologies, and the best accuracy is found when using homogeneous training databases highly similar to the target testing structure.  
	However, heterogeneous training databases provide a just slightly less good overall prediction performance, while the trained kernels display a much higher degree of versatility, predicting accurate forces also in out-of-sample tests.
	Furthermore, 3-body ML-FFs trained on an heterogeneous database accurately reproduce structural fingerprints such as pair distance distribution functions at finite temperatures.
	Using this ML-FF to derive a non parametric machine-learning mapped force field (``M-FF") via the procedure discussed in Ref.~[\onlinecite{Glielmo2018}]  makes it possible to execute tens of ns long simulations with a minimal computational effort, allowing to assess the thermal stability of Ni$_{19}$ nanoclusters.

	In the next section, we introduce the necessary GP formalism (II A), provide expressions for the kernels used throughout the work (II B) and briefly explain the concept of ``mapped" force field \cite{Glielmo2018} (II C).
	We then describe a protocol for the validation of force predictions (III A) and discuss the performance achieved by the three kernels when tested on structures   either very similar or morphologically different from the ones present in the training database for single-structure (III B) and multi-structure (III C) training. 
    The construction of the 3-body M-FF and its   validation are described (III D), while its application in MD simulations investigating the behaviour of Ni$_{19}$ clusters in the 300-1200~K temperature range is described in section III E.
	
\section{Machine Learning \\ force fields}

\subsection{Gaussian process regression}

A GP regression \cite{rasmussen2006gaussian} is a Bayesian method to learn a function from a finite database $\mathcal{D}$ of input-output pairs. 
As we are interested in learning the local force acting on any given atom, we construct such training database by extracting (from a DFT simulation) a set of local configurations $\rho_i$ relative to each atom and the corresponding forces $\mathbf{f}_i$ on that atom. 
This database $\mathcal{D} = \{( \rho_i, \mathbf{f}_i )\}_{i=1}^N$ is then partitioned into a training set $\mathcal{D}_{tr}$ (with $N_{tr}$ entries) used for learning, and a test set $\mathcal{D}_{test}$ (with $N_{test}$ entries) used for validation.
As the space of forces is three dimensional, we here use the multi-output (vectorial) version of GP regression\cite{alvarez2012kernels, micchelli2005kernels, Glielmo:2017dj}, for which the learned function $\mathbf{f}(\rho)$ takes the form
\begin{equation}
\mathbf{f}(\rho) = \sum_{d=1}^{N_{tr}} \mathbf{K}(\rho,\rho_d) \boldsymbol{\alpha}_d,
\label{eq:GP_def}
\end{equation}
where $\mathbf{K}$ is a matrix-valued kernel function encoding the correlation of the forces relative to any two atomic environments. 

The coefficients $\boldsymbol{\alpha}_i$ in Eq.~\eqref{eq:GP_def} can be written in closed form as 
\begin{equation}
\boldsymbol{\alpha}_d = \sum_{d'}^{N_{tr}} [\mathbb{K}+\mathbb{I}\lambda]^{-1}_{dd'}\cdot \mathbf{f}_{d'},
\label{eq:alpha_def}
\end{equation}
where $\mathbb{K}$ is the covariance matrix  containing $N_{tr} \times N_{tr}$ block entries $\mathbb{K}_{dd'} = \mathbf{K}(\rho_d, \rho_{d'})$, $\mathbb{I}$ is the identity matrix and $\lambda$ is a  regularization hyperparameter that formally governs the uncertainty associated with the training dataset outputs , which has been kept fixed at a value of 10$^{-5}$.
The performance of a GP is determined by the choice of the kernel function $\mathbf{K}$, its hyperparameters and by the choice of training set $\mathcal{D}_{tr}$.

\subsection{Kernel functions}

The kernel function should be invariant w.r.t. translation and permutation of identical atoms, and covariant (when predicting forces) w.r.t. rotation of the configurations. 
Furthermore, the function $\mathbf{K}$ should have a spatial resolution compatible with the features of the energy landscape encoded in the training dataset; this is taken care of by optimizing the kernel hyperparameters.
A useful property of a kernel function for force or energy prediction is its $\mathit{order}$, that is, the maximum number of simultaneously interacting particles it can describe (see  
Ref.[\onlinecite{Glielmo2018}] for a formal definition). 
Here we will use 2-body, 3-body and many-body force kernels \cite{Deringer:2017ea, Glielmo:2017dj}.\\

\paragraph{2-body.}
We assume that the force $\mathbf{f}$ acting on an atom located at position $\mathbf{r}_a$ is a sum of independent contributions associated with every other atom in its local environment $\rho_a$. 

Each configuration is expressed as a sum of Gaussian functions of width $\sigma$ representing individual atoms ``SOAP representation" \cite{Bartok:2013cs}).
A natural way to obtain a rotation invariant scalar energy kernel would be via integration over the group of rotations of the space-integrated overlap of each couple of configurations \cite{Bartok:2013cs}.  
The covariance of predicted forces, a general property not requiring the existence of an underlying invariant total energy function, can also be obtained as an integral over rotations from a suitable matrix-valued covariant integration expression, as detailed in Ref.[\onlinecite{Glielmo:2017dj}].  
For the 2-body kernel case the integration can be carried out analytically, yielding the following matrix-valued energy-conserving kernel\cite{Glielmo:2017dj}:

\begin{eqnarray}\label{eq:2body_rot}
\mathbf{K}_{2}^{s}(\rho_a,\rho_b) & = & \sum_{\substack{i\in\rho\\j\in\rho}} \phi(r_{ai},r_{bj}) \mathbf{\hat{r}}_{ai}  \mathbf{\hat{r}}_{bj}^{T},\\
\phi(r_{ai},r_{bj}) & = & \frac{\mathrm{e}^{-\alpha_{ij}}}{\gamma_{ij}^{2}}(\gamma_{ij}\cosh\gamma_{ij}-\sinh\gamma_{ij}),\nonumber \\
\alpha_{ij} & = & \frac{r_{ai}^{2}+r_{bj}^{2}}{4\sigma^{2}},\nonumber \\
\gamma_{ij} & = & \frac{r_{ai}r_{bj}}{2\sigma^{2}}, \nonumber 
\end{eqnarray}
where $\mathbf{r}_{ai}$ expresses the position relative to the central atom
of its $i_{th}$ neighbour.  
\\

\paragraph{3-body.}
A 3-body kernel allows to represent an angular dependence on the force components. 
As described in detail in Refs.[\onlinecite{Glielmo2018}, \onlinecite{Bartok:2015iw}], a 3-body force kernel can be built as a double derivative of a 3-body energy kernel with respect to the positions of the central atoms of the configurations $\rho_a$ and $\rho_b$:
\begin{equation}
\mathbf{K}^{s}_{3} (\rho_a, \rho_b) = \dfrac{\partial ^2k ^{s}_{3} (\rho_a, \rho_b ) }{\partial \mathbf{r}_a \partial \mathbf{r}^T_b}.
\label{eq:force_kernel_3b}
\end{equation}

The 3-body energy kernel $k ^{s}_{3}$ compares triplets of atoms that include the central atom across the two configurations. 
This kernel is intrinsically invariant under permutation, rotation and translation of the atoms in $\rho_a$ and $\rho_b$, avoiding the need of any integration over SO(3).
Each triplet is associated with a vector $\mathbf{q}_{aij}$ containing the three atomic distances i.e., $\mathbf{q}_{aij} = (r_{ai},r_{aj},r_{ij})^T$. 
Apart from a normalisation factor, the 3-body kernel reads: 
\begin{equation}
k_3^s(\rho_a, \rho_b)	 =\sum_{\substack{i,j \in \rho_a \\
		k,l \in \rho_b}} \sum_{\mathbf{P} \in \mathcal{P}_c} \mathrm{e}^{-\|\mathbf{q}_{aij}- \mathbf{P} \mathbf{q}_{bkl}\|^2/2\sigma^2},
\label{eq:energy3_kernel}
\end{equation}
where $\mathcal{P}_c$  ($|\mathcal{P}_c| = 3$) is the set of cyclic permutations of three elements, 
and  $\sigma$ is the single required lengthscale hyperparameter.
Summing over the permutation group is needed to guarantee permutation symmetry of the energy.
 No $i\neq j$ or $k\neq l$  restriction is however imposed in the external sum, making the overall expression not limited to the case of three distinct atoms, so that  the kernel also includes the 2-body case as a subset.  
We note that permutation invariance in 3-body kernels could also be obtained using permutation invariant descriptors, as done in Ref.[\onlinecite{Deringer:2017ea}]\\

\paragraph{Many-body.}
Describing arbitrarily complex interactions requires a many-body kernel function such that force prediction becomes dependent on the full local atomic environment $\rho_a$, and is no more the result of summing independent pairwise (or triplet) contributions. 
A way to obtain a many body kernel $k_{MB}$ (see Ref.[\onlinecite{Glielmo2018}]) is to take the exponential of a scalar 2-body kernel $k_2$:
\begin{equation}
k_{MB}(\rho_a, \rho_b) = \mathrm{e}^{  k_2 (\rho_a,\rho_b )/ \theta^2 },
\label{eq:mb_kern}
\end{equation}
where
\begin{equation}
k_{2}(\rho_a,\rho_b)  = \sum_{\substack{i \in \rho_a \\
		j \in \rho_b}} \mathrm{e}^{-\|\mathbf{r}_{ai}- \mathbf{r}_{bj} \|^2/2\sigma^2}.
\label{eq:2b_nonrot_kern}
\end{equation}
To impose rotational force covariance we should perform an integration over the $SO(3)$ manifold of rotations \citep{Glielmo:2017dj}.
Unfortunately  the integration over SO(3) of the many-body kernel in Eq.(\ref{eq:mb_kern}) cannot be done analytically, while numeric integration is computationally heavy. 
We hence resort to restricting the summation to a discrete symmetry group $R$ of rotations (and reflections) whose elements $\mathcal{R}$ are represented by orthogonal matrices $\mathbf{R}$:
\begin{equation}\label{eq:mb_cov_kern}
\mathbf{K}_{MB}^{ds}(\rho_a, \rho_b) =   \dfrac{1}{|R|} \sum_{\mathcal{R} \in R} \hspace{0.1cm} \mathbf{R} \hspace{0.1cm}  k_{MB}(\rho_a, \mathcal{R} \rho_b).
\end{equation}
The optimal choice of rotation group is system-dependent: in FCC and BCC bulk environments a natural choice is to sum over all elements of  the O$_{h}$ point group.
The resulting many-body kernel can be made arbitrarily accurate if given a large enough training set \cite{Glielmo2018, csaji2001approximation} while the predicted force field will not  be conservative (make zero work on closed loops to numerical accuracy) by construction. 
However, to the extent that force errors are small, the energy-conserving character of the reference Hamiltonian forces will be approximately reproduced. 

\subsection{Mapped force field (M-FF)}
Once the 3-body GP has been trained, the ``mapping" technique described in Ref.[\onlinecite{Glielmo2018}] can be used to build a non-parametric 3-body force field (a M-FF) which retains the accuracy of the original GP while typically increasing its computational speed by a factor $10^3 - 10^4$. 
This procedure is effectively equivalent to storing the energies predicted by the kernel (\ref{eq:energy3_kernel}) for a three-dimensional grid of values of the triplet of distances ($r_{ai}$, $r_{aj}$, $r_{ij}$) occurring in a three atom system.
In a more complex structure, the contributions from every triplet and atom pair are calculated by spline interpolation over   the stored GP predictions of the energy values.  
Analytic differentiation of the spline expression produces an energy conserving force field practically indistinguishable from the predictions of the 3-body GP used to build it, while independent of the 
number of configurations $N_{tr}$ used for GP training. 
This M-FF could be seen as a classical $n$-body force field, as simple to physically interpret and fast to compute as a standard parametrised 3-body force field, whose systematic non-parametric construction requires no \emph{ad-hoc} parameter choice and fine-tuning. 
This enables simulation times which would not be achievable by standard direct GP force prediction or by first-principles molecular dynamics based on the reference DFT Hamiltonian. 

\section{Results}

\begin{figure}[t]
	\includegraphics[width=1\linewidth]{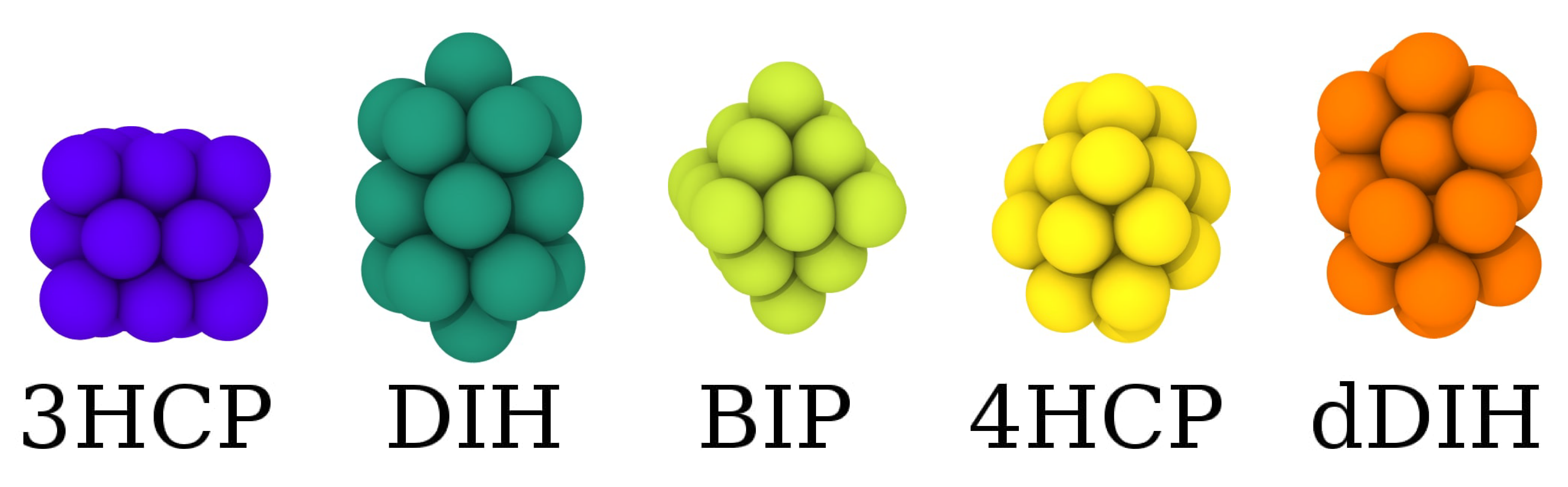}

	\caption{Five Ni 19 structures ordered, from left to right, according to an energetic criterion, with the first one being the global minimum.}
	\label{fig:clusters}
\end{figure}

We consider Ni nanoclusters of 19 atoms and gather data from Born-Oppenheimer molecular dynamics (BOMD) simulations at 300K, 600K, and 900K, from five different initial structures, represented in Figure \ref{fig:clusters}: a hcp motif of three layers (3HCP), a double icosahedron (DIH), a bipyramid (BIP), a four stacked hcp layer (4HCP) structure, and a structure obtained by displacing two five-fold arranged vertexes of a double icosahedron to form two six-fold rings, also known as a ``rosette" defect \cite{apra2004amorphization} (dDIH).
The first three motifs are the most energetically favourable at PBE DFT level \cite{lu2011structural}, the other two were found with a metadynamics sampling procedure\cite{pavan2013sampling, santarossa2010free} and are included here to introduce low symmetry morphologies in the database set. 
The BOMD simulation details are provided in the supplementary material.

\subsection{Validation methodology} 
It is evident from Eq. (\ref{eq:GP_def}) and (\ref{eq:alpha_def}) that the predictions of a GP will depend on how the training dataset $\mathcal{D}_{tr}$ is chosen. 
To assess the errors incurred by the three kernels while used in interpolation and (putative) extrapolation regimes, we next systematically analyse the GP predictions on test databases containing contributions from all five structures while training is carried out on different combinations of structures.
This procedure allows us to introduce a novel strategy to measure the similarity between cluster geometries, based on the relative GP errors made while training and testing on two different structures.
In our practical implementation, for all the GP trainings, we choose $N_{test}$ = 400 for each of our five cluster structures, yielding a total pool of 2000 testing points.

Every test is repeated five times to estimate a standard deviation for the Mean Absolute Error on Forces (MAEF),  defined as the average error done by the GP on the force vector:
\begin{equation}
	\text{MAEF} = \dfrac{1}{ N_{test}} \sum_{d=1}^{N_{test}} \sqrt{ \sum_{c=1}^3 \left( \text{f}_{d,c} - \text{f}^{\,\prime}_{d,c} \right)^2},
	\label{eq:MAE}
\end{equation}
where $\mathbf{f}$ and $\mathbf{f'}$ are the reference and predicted forces acting on an atom, respectively, and $c$ indicates the Cartesian component.
Our tests can be separated into three categories: \textit{self-training}, \textit{cross-training} and \textit{mixed-training}, depending on which databases was used to build the training sets and which subset of the testing pool is used.  
In the \textit{self-training}, the configurations used to build the $\mathcal{D}_{tr}$ and $\mathcal{D}_{test}$ are associated with the same cluster structure.  
In the \textit{cross-training}, the database from a structure morphology is used for training, and testing occurs on configurations associated with the other four structures. 
In both of the above, the database is homogeneous, that is it contains configurations relative to a single morphology.
For the \textit{mixed-training} we build and test all possible heterogeneous training sets $\mathcal{D}_{tr}$ that contain inputs from two, three, and all five morphologies (here as in \textit{self-training}, no data point present in $\mathcal{D}_{tr}$ is allowed to be in $\mathcal{D}_{test}$). 

\subsection{Self- and cross-training} 

\begin{figure}[t]
	\includegraphics[trim = 0.5cm 0.5cm 1cm 0cm, clip, width=1\linewidth]{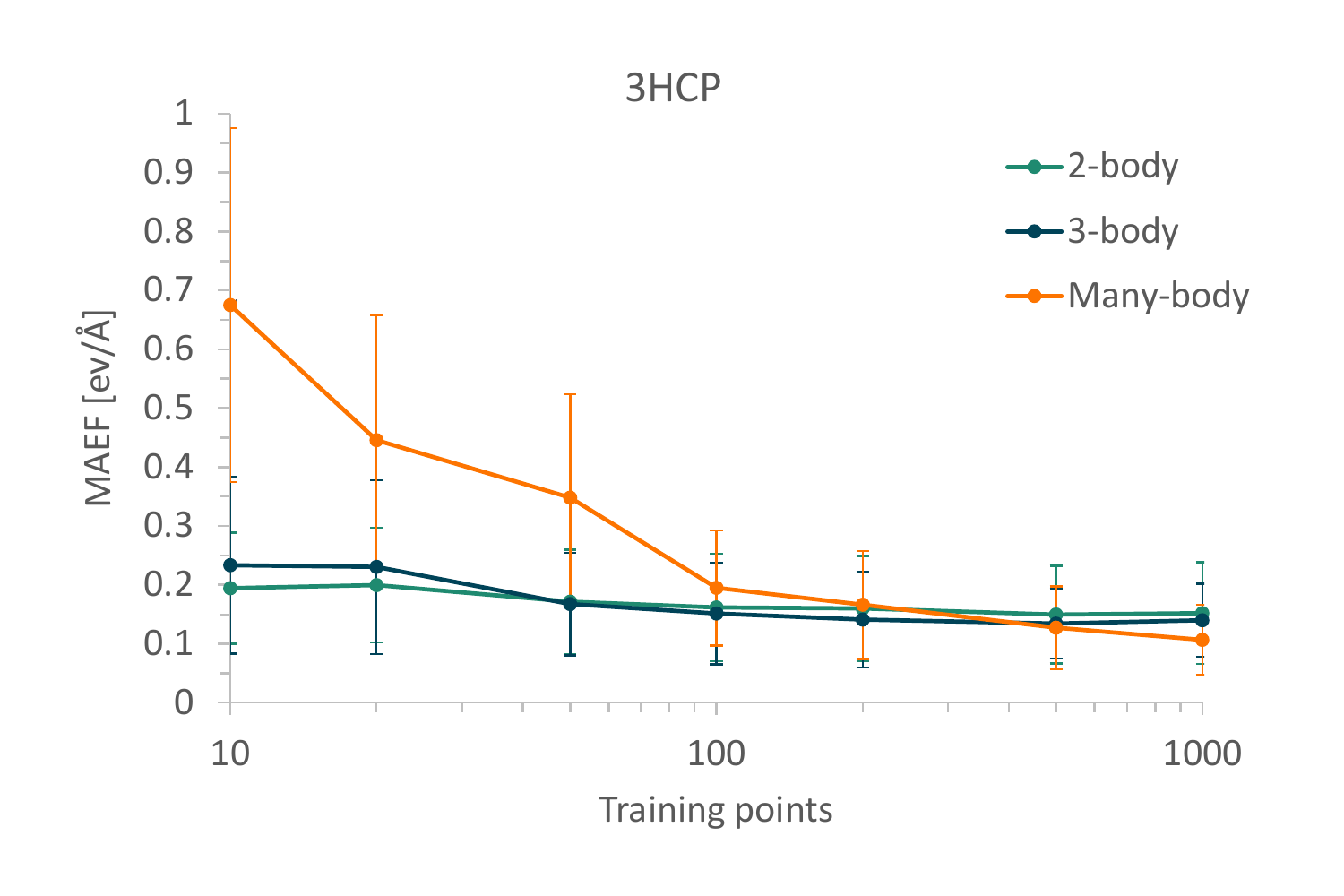}
	\caption{Learning curves for the paradigmatic example of training and testing on a 3HCP morphology. The k	ernel $n$-body order is colour coded, with 2-body represented in cyan, 3-body in blue, and many-body in orange.}
	\label{fig:learning_curves}
\end{figure}
\begin{figure}[t]
	\includegraphics[trim = 1cm 1cm 1cm 0cm, clip, width=1\linewidth]{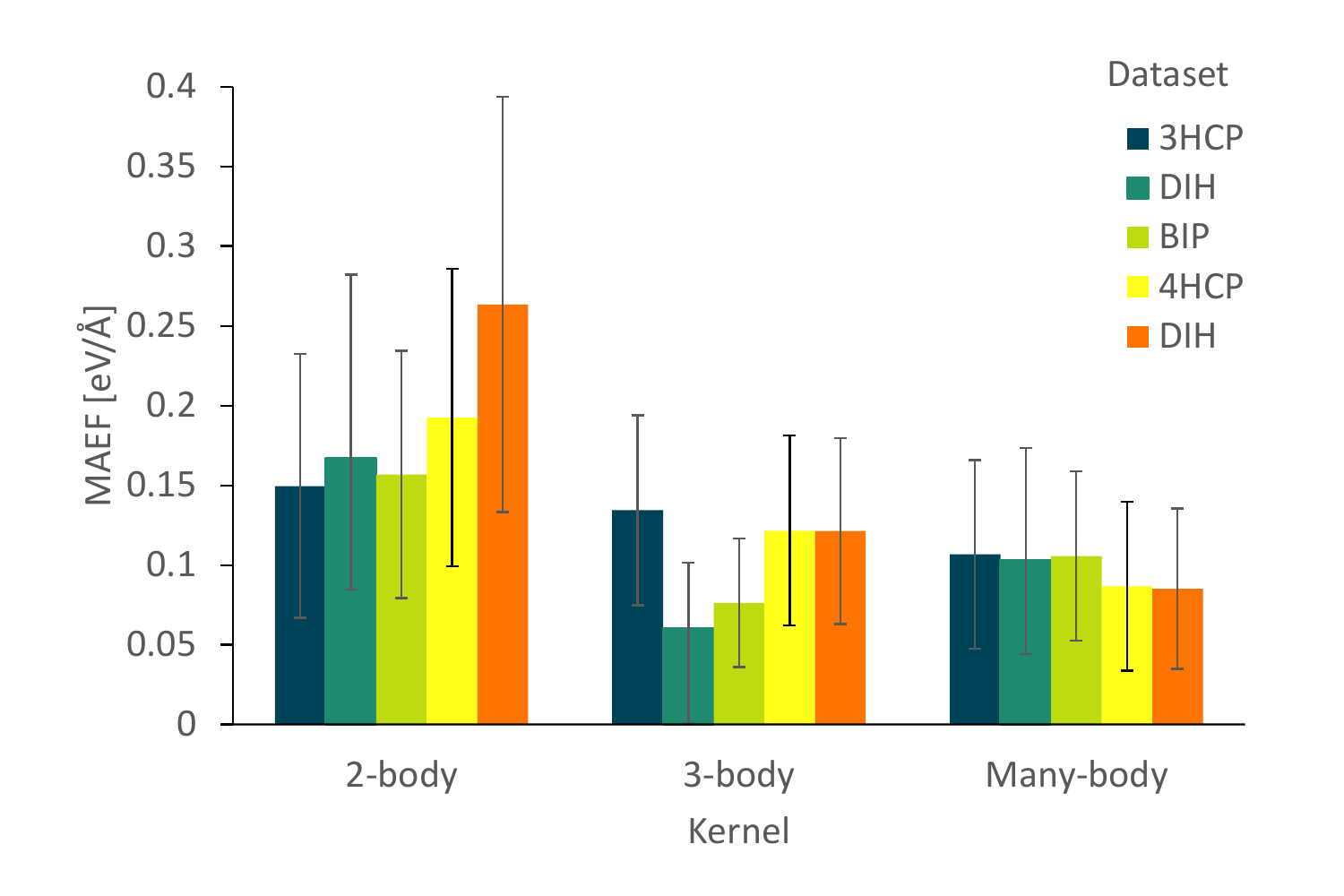}
	\caption{MAEF for GP force prediction for the 5 clusters in Figure \ref{fig:clusters} and for 3 kernels, with standard deviation obtained from 5 tests.
	For all tests, $N_{test}$ = 400 and $N_{tr}$ = 500 (2- and 3-body kernels) or  $N_{tr}$ = 1000 (many-body kernel).}
	\label{fig:converged_MAEF}
\end{figure}

We first discuss the results for \textit{self-training}.
Figure \ref{fig:learning_curves} reports a learning curve (MAEF against $N_{tr}$) for the case of a 3HCP cluster structure.
The 2-body kernel achieves its maximum accuracy for $N_{tr}$ $>$ 50; similarly, the 3-body MAEF decreases with $N_{tr}$ until $N_{tr}$ $ > $ 100 and an accuracy plateau is reached.   
The accuracy of the many-body kernel, on the other hand, keeps increasing with the number of training set point, as expected for an universal approximator kernel \cite{csaji2001approximation, Glielmo2018}.
The learning curves for the other structures show the same qualitative trends (see supplementary material).
Figure \ref{fig:converged_MAEF} shows the converged MAEF achieved by \textit{self-training} GPs for each of the five morphologies when using 2-, 3-, and many-body kernels.
The left-hand histogram reveals that modeling the atomic interactions between Ni atoms in terms of a 2-body potential yields a MAEF larger than the target accuracy of  0.15 eV/$\text{\AA}$ for all morphologies, with higher values for low symmetry ones (4HCP and dDIH). 
We note that this is not the case for FCC bulk Ni systems, where 2-body kernels were found to be suprisingly accurate \cite{Glielmo:2017dj}.  
The central and right-hand histograms in Fig.~\ref{fig:converged_MAEF} reveal that both 3- and many-body kernels achieve a suitably accurate force prediction for all cluster structures if the training dataset used contains 200+ points.
For comparison, the calculated MAEF of a state-of-the-art classical parametric potential for Ni  \cite{purja2009development} is  0.59 $\pm$ 0.39 eV/$\text{\AA}$. 
The relative importance of $n$-body contributions to the forces in the five Ni$_{19}$ cluster morphologies can be appreciated by looking at the accuracy of the $n$-body kernels. 
For instance, the accuracy of 2-body and 3-body forces is very similar for the 3HCP morphology, indicating that the angular dependence of forces is not crucial in this motif, while it is more significant for the other structures.
We note that comparing $n$-body kernel predictions could be more generally used as a way to characterize the nature of the bonding occurring in complex systems such as metal nanoclusters or grain boundaries, and to reveal and quantify (dis)similarities between these systems or relative to reference bulk structures. 

\begin{figure}[t!]
	\includegraphics[width=1\linewidth]{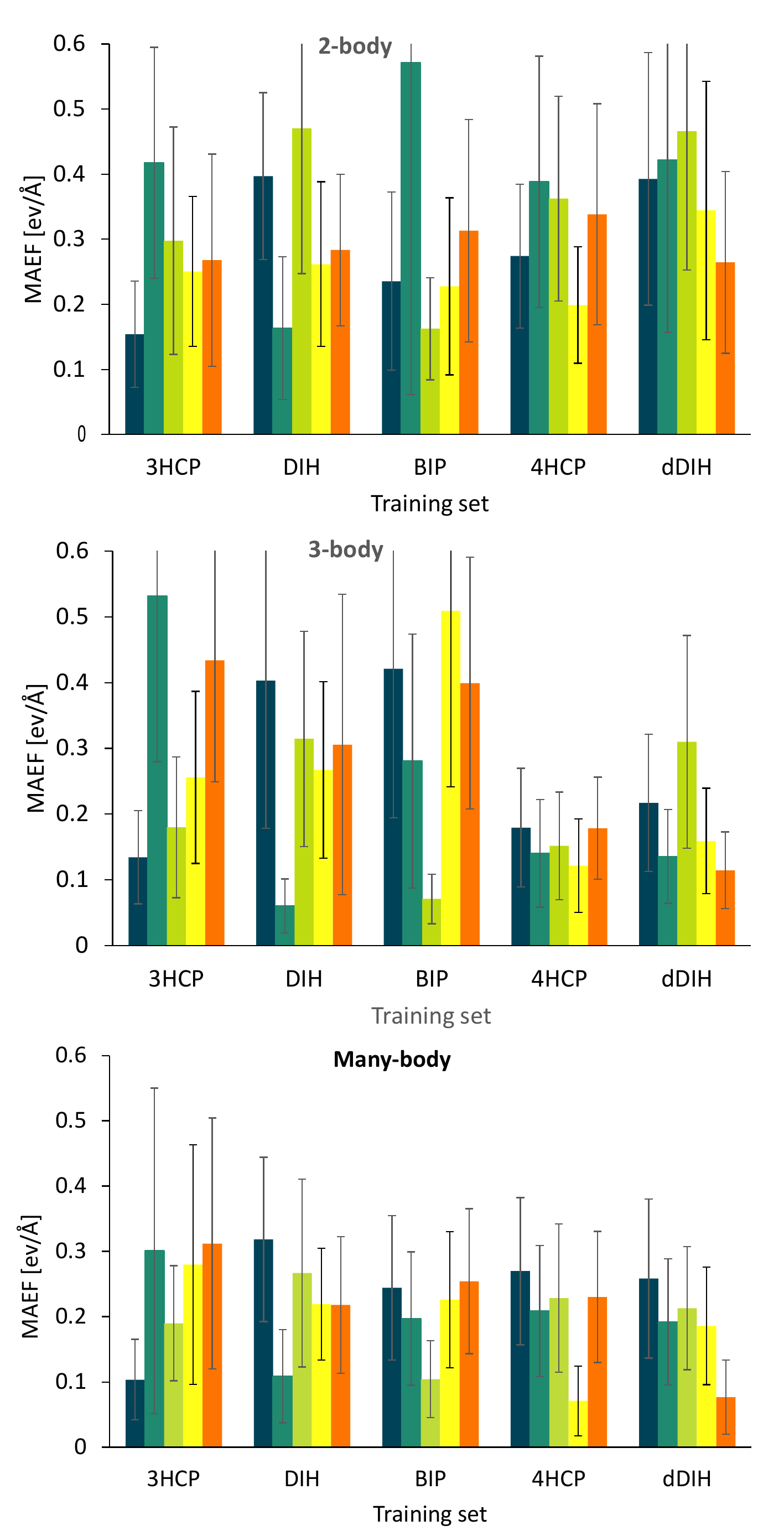}

	\caption{From top to bottom we report the MAEF and its standard deviation for the 2- ($N_{tr} = 500$), 3- ($N_{tr}$ = 500) and the many- ($N_{tr}$ = 1000) body kernels trained and tested on different Ni-19 datasets at 300~K.}
	\label{fig:ni_cross_err}
\end{figure}
The MAEFs obtained for \textit{cross-testing} are reported in Figure \ref{fig:ni_cross_err}.   
In these case, the 2-body kernels MAEFs are consistently larger than  0.15~eV/$\text{\AA}$ and often twice as large.
Comparing the 3- and many-body kernels reveals the accuracy achieved by the 3-body kernel strongly depends on the training database, while the many-body kernel displays more consistent errors over different structures. 
This could be rationalised by considering that a many-body kernel is capable of learning high-$n$ interaction terms whose contributions are effectively sampled in any morphology, even e.g., in structures in which they are quantitatively less important. 
These terms help maintaining a good prediction accuracy even on ``partially unknown" new morphologies where higher order interactions come more into effect.
The 3-body kernel is instead intrinsically restricted to 3-body interactions, and
if the reference forces include (say) a 4-body interaction contribution, incorporating this by machine learning based on a lower-dimensional feature space may achieve some success only in \textit{self-training} (interpolation) mode, but won't correctly extrapolate to new structures.
This suggests that the accuracy that a 3-body kernel achieves on a target structure is to a significant extent conditional to the presence of database entries representative of that structure in the training database used.
\begin{figure}[t!]
\includegraphics[width=1.0\linewidth]{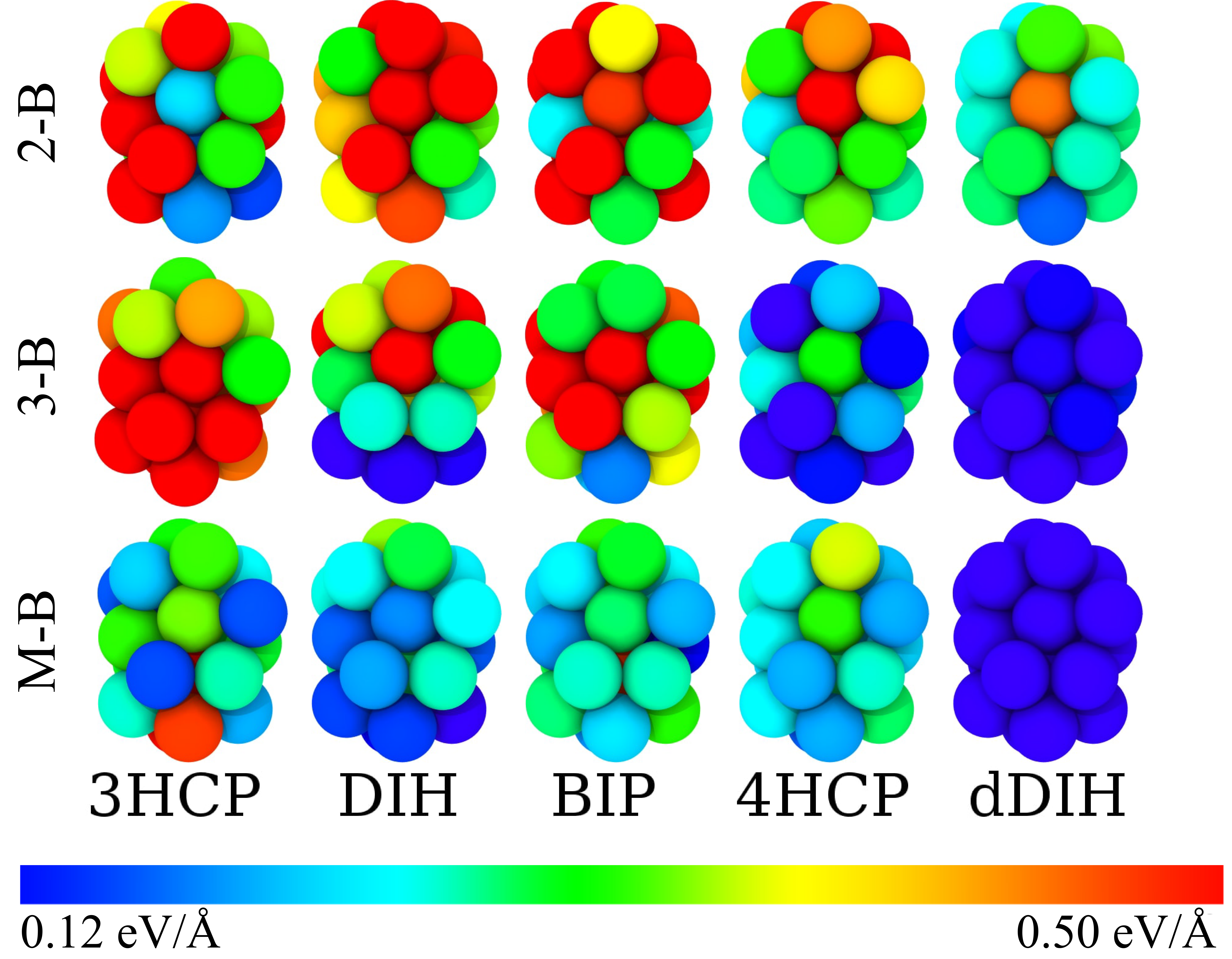}
	
	\caption{MAEF incurred by  the 2-(top), 3-(central), and many-body (bottom) kernels on each dDIH atoms when trained on (left to right): 3HCP, DIH, BIP, 4HCP, dDIH. Color coding ranges from 0.12 eV/$\text{\AA}$ (blue) to 0.5 eV/$\text{\AA}$ (red).}
	\label{fig:error_on_dDIH}
\end{figure}
Consistently, for this kernel, the training databases comprising the low-symmetry morphologies (4HCP and dDIH) that have the most varied repertoires of atomic environments are those which work best in \textit{cross-testing}. 
We also note how the \textit{cross-testing} error incurred by training over 4HCP and testing on its higher-symmetry counterpart 3HCP is 0.18~eV/$\text{\AA}$, while the reversed test yields a significantly larger  0.26~eV/$\text{\AA}$ MAEF.  
The same effect becomes even more apparent by examining the dDIH (low symmetry) and DIH (high symmetry) pair of structures, yielding  0.14~eV/$\text{\AA}$ and  0.31~eV/$\text{\AA}$ errors in the direct and reversed tests, respectively.

Further analysis of the GP predictions allows some qualitative understanding of why using different training databases leads to stronger or weaker performances over the available testing sets. 
We first examine the case of training on each of the five structures and testing on the dDIH structure. 
Figure \ref{fig:error_on_dDIH} is a visual representation of the MAEFs committed at the testing stage by our three kernels after they were trained on the five single-structure training databases.  
As expected from Figure \ref{fig:ni_cross_err}, the 2-body kernel is associated with large errors for all training sets but the dDIH one - the only one here in the \textit{self-training} regime. 
In the case of 3-body kernel, training on a 4HCP database yields the best overall \textit{cross-training} performance (while as expected, \textit{self-training} on a dDIH database offers better results). 
This provides good accuracy on most atoms, falling short only around the rosette defect, a peculiar distortion absent in the 4HCP structure. 
The MAEFs incurred by training on the DIH database are also very low for the lowermost 5-fold cap of the dDIH cluster. 
This is expected since these local environments are very similar in these two morphologies. 
On the other hand, \textit{cross-training} on the BIP and 3HCP datasets fails to predict forces around the icosahedral centres and the rosette defect of the dDIH. 
These results hold true even in the case of the many-body kernel, 
for which the DIH is the best performing \textit{cross-training} morphology. 

\begin{figure}[t!]
\includegraphics[width=1\linewidth]{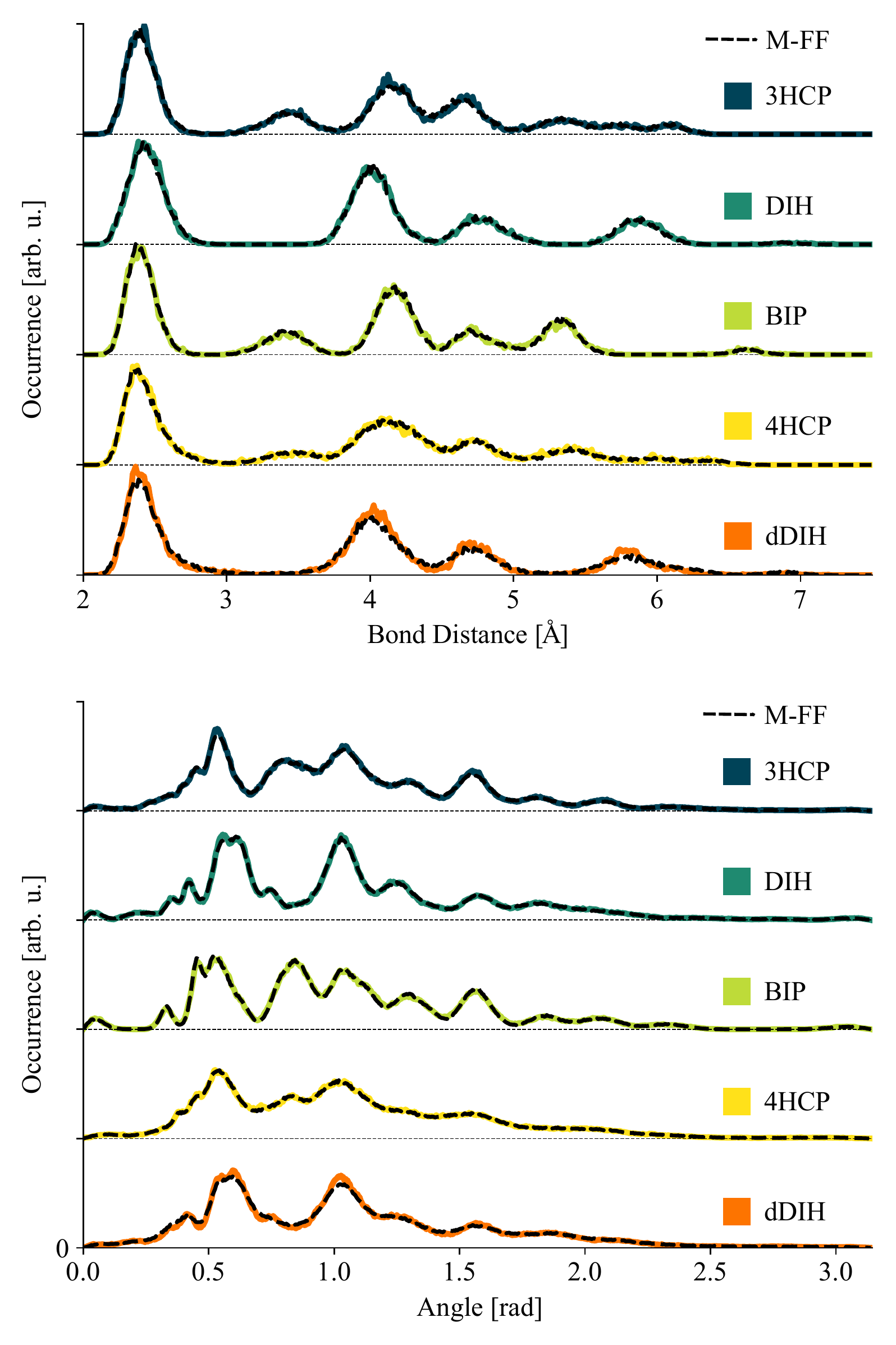}
	\caption{Pair-distance (top) and bond angle (bottom) distribution functions for the five cluster structures,  averaged over 2~ps from 300~K BOMD simulations (in colour) and M-FF MD simulations (black dashed).}
	\label{fig:pdf}
\end{figure}

We next compare the pair-distance function (PDF) and the bond angular distribution function (BADF) for the five morphologies as obtained from BOMD simulations at 300~K, reported in Figure \ref{fig:pdf}.
These reveal structural differences between the morphologies. 
For instance,  the PDF peak close to 3.3~$\text{\AA}$ in the 3HCP, 4HCP, and BIP morphologies is absent in the DIH and dDIH structures.
Also, the BADF in the bottom panel displays a broadened distribution for 4HCP and dDIH and much more peaked ones for 3HCP, DIH, and BIP.

\begin{figure}[t!]
\includegraphics[width=1\linewidth]{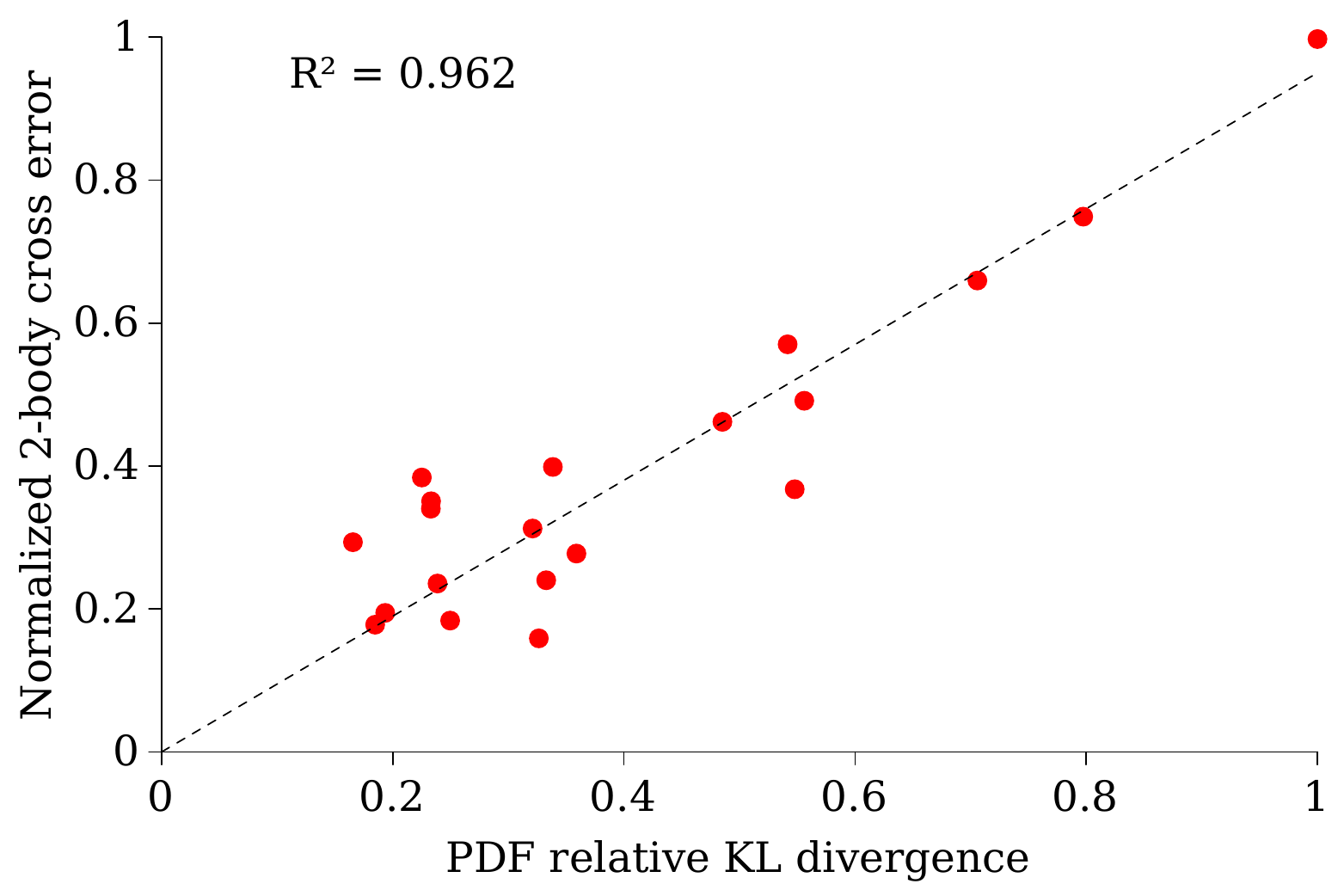}
	\caption{Normalised scatter plot highlighting the correlation between the KL divergence calculated on ordered pairs of PDFs from Figure \ref{fig:pdf} and the \textit{cross-training} MAEFs incurred for the corresponding pairs of cluster structures (see text for more details).}
	\label{fig:KLD}
\end{figure}

As a possible quantitative indicator of how well a PDF ``samples" another one we calculate their Kullback$-$Leibler (KL) divergence.
For a discrete probability distribution this is calculated as:
\begin{equation}
	KL (P || Q) = \sum_i P(i) \log \dfrac{P(i)}{Q(i)}.
	\label{eq:KL}
\end{equation}
This (asymmetric) quantity measures the information ``lost" when a function $Q$ is used to approximate another function $P$, returning a 0 value when $P$ = $Q$, and increasing positive values as $P$ grows dissimilar from $Q$.  
In the present context, $Q$ and $P$ are the PDFs associated with the training set and testing set, respectively. 
Figure \ref{fig:KLD} contains a normalized scatter plot 
comparing the KL divergence relative to each ordered pair of PDFs taken from Figure \ref{fig:pdf} with the corresponding  \emph{cross-training} MAEF incurred by the 2-body kernel (see supplementary material for details on how these two quantities were normalized). 
 The graph reveals a striking correlation between the two dissimilarity measures, generally highlighting the importance of the presence of database entries which contain pairs of atoms at distance values relevant for the testing dataset. 
Moreover, since the PDF can be assumed to be an unique structural descriptor in the case of monometallic nanoparticles \cite{tribello2011exploring, oganov2009quantify, steenbergen2014two, pavan2015metallic}, the correlation indicates that the 2-body cross-training error is non-trivially linked to properties that go beyond 2-body descriptors.
Thus, the KL divergence between PDFs could be used as an \emph{a priori} indicator of extrapolating performance of the 2-body kernel in \textit{cross-training}. 
Similar tests for the 3-body kernel also display a positive correlation, although of smaller statistical significance (see supplementary material).
The results above suggest that evaluating the KL divergence for other functions than the PDFs could provide more dissimilarity estimators. 
This could be used to guide the extraction of informed, minimally sized training databases from a  ``general" database too large to be used in full for GP regression.  


\subsection{Heterogeneous training and training set optimisation} 

\begin{figure}[t]
\includegraphics[width=1\linewidth]{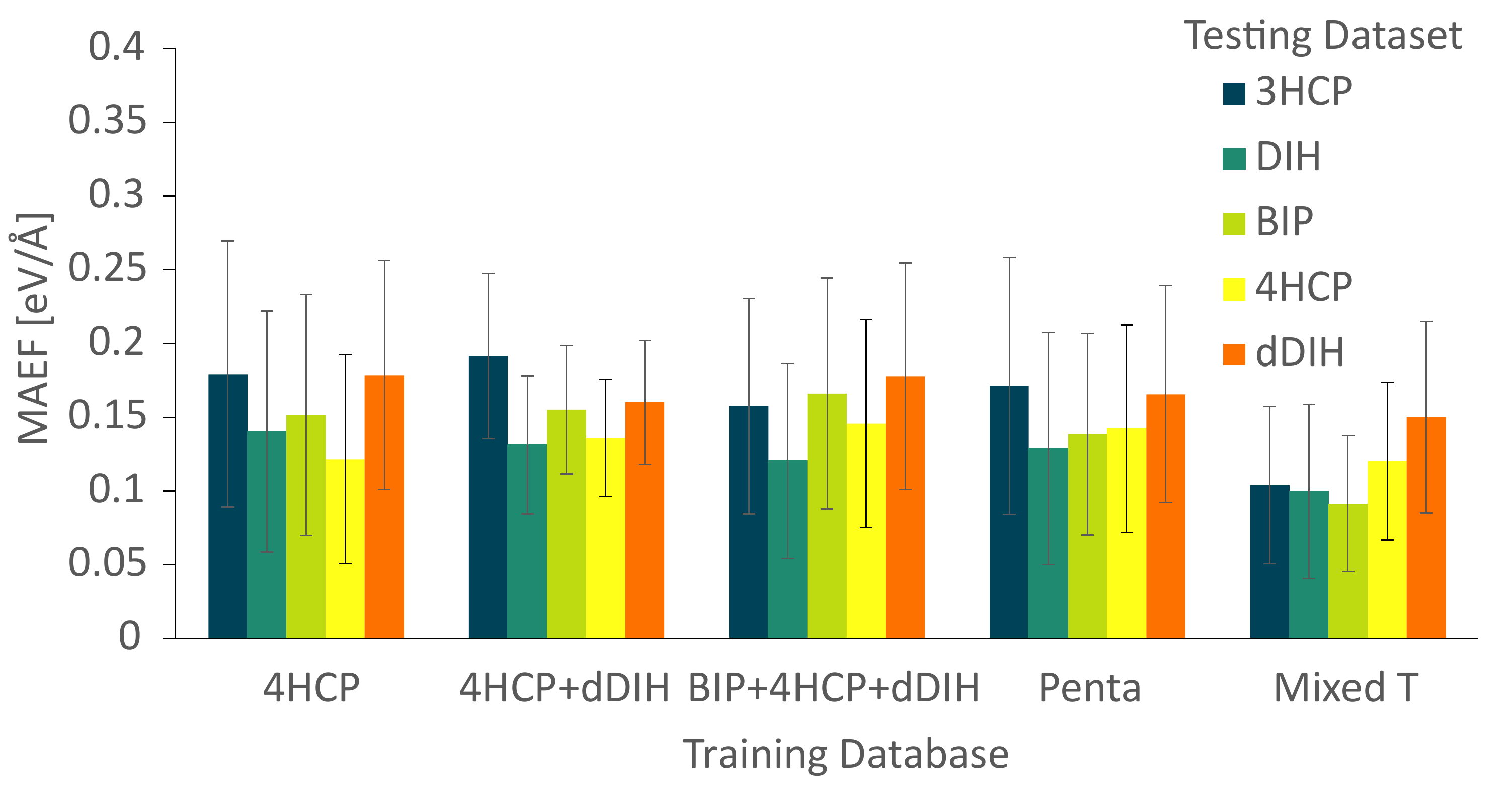}
\caption{MAEFs for four 3-body kernels trained on 300~K databases containing 500 configurations  representative of one, two, three and all five cluster morphologies,  and a ``mixed T" database of 1000 DFT configurations extracted from 300, 600  and 900~K simulations including all cluster morphologies.}
	\label{fig:best_3B_ker}
\end{figure}

We next examine the \textit{mixed-training} strategy, to learn how a small training database $\mathcal{D}_{tr}$ which still encompasses a sufficiently varied repertoire of atomic distances and bond angle values could be built. 
For this, we test the accuracy of training on mixed datasets containing entries from two, three, and all five different cluster morphologies. 
Our results indicate that the MAEF incurred by the 3-body kernel is fairly homogeneous in the various \textit{mixed-training} and testing scenarios, staying the same within a  0.03~eV/$\text{\AA}$ standard deviation, contrary to what was generally found for \textit{self-training}.  
This suggests that introducing even a modest amount of variety in the training configuration pool is sufficient to achieve a reasonably complete training of a 3-body kernel,  avoiding ``local'' overfitting causing extrapolation errors.  
Consistently, the MAEF incurred by a kernel not restricted to just a 3-dimensional feature space and thus much harder to completely train, such as the many-body kernel, is found to have a higher standard deviation ( 0.05~eV/$\text{\AA}$) when trained and tested in the same scenarios. 
For all kernels, we find that \textit{mixed-training} yields errors comprised between those incurred by \textit{self-training} and \textit{cross-training}, with MAEFs slightly higher than those produced by \textit{self-training} but appreciably smaller than those associated with \textit{cross-training}. 

Figure \ref{fig:best_3B_ker} illustrates the performance of the 3-body kernel trained on our ``best" single-structure database (4HCP, see section III B), the best choice of two- and three-structure mixed databases, the full 5-structure database  (``penta"), and a (``mixed T") training database containing 1000 DFT configurations extracted from simulations at 300, 600 and 900~K. 
The results are good for all training scenarios and notably, the 5-structure ``penta" databases achieves the same performance of all the other database choices which needed to be identified as the best restricted ones. 
To investigate the performance stability of a given, simple database construction recipe, we generated 100 independent  ``penta" $\mathcal{D}_{tr}$ training sets, each containing 100 randomly chosen configurations for each cluster structure. 
Training 3-body kernels on  the low temperature $\mathcal{D}_{tr}$ of Figure \ref{fig:best_3B_ker} and testing on a fixed database also comprising configurations from all five structures yields an average MAEF of  0.14 $\pm$  0.07 eV/$\text{\AA}$. 
The small  0.004 eV/$\text{\AA}$ difference we find between the MAEF incurred by the best-performing  low-temperature GP and the average MAEF suggests that the accuracy gain which might be obtained by a ``best training set choice" procedure is practically negligible. 
To further analyse this issue we performed Metropolis Monte Carlo simulations to optimize the dataset training points, and again found no significant accuracy gain (see supplementary material). 
 An overall better performance was instead achieved when using the ``mixed T" database which included higher T configurations for all cluster morphologies.

\subsection{Building and validating a 3-body M-FF} 

\begin{figure}[t]
\includegraphics[width=1\linewidth]{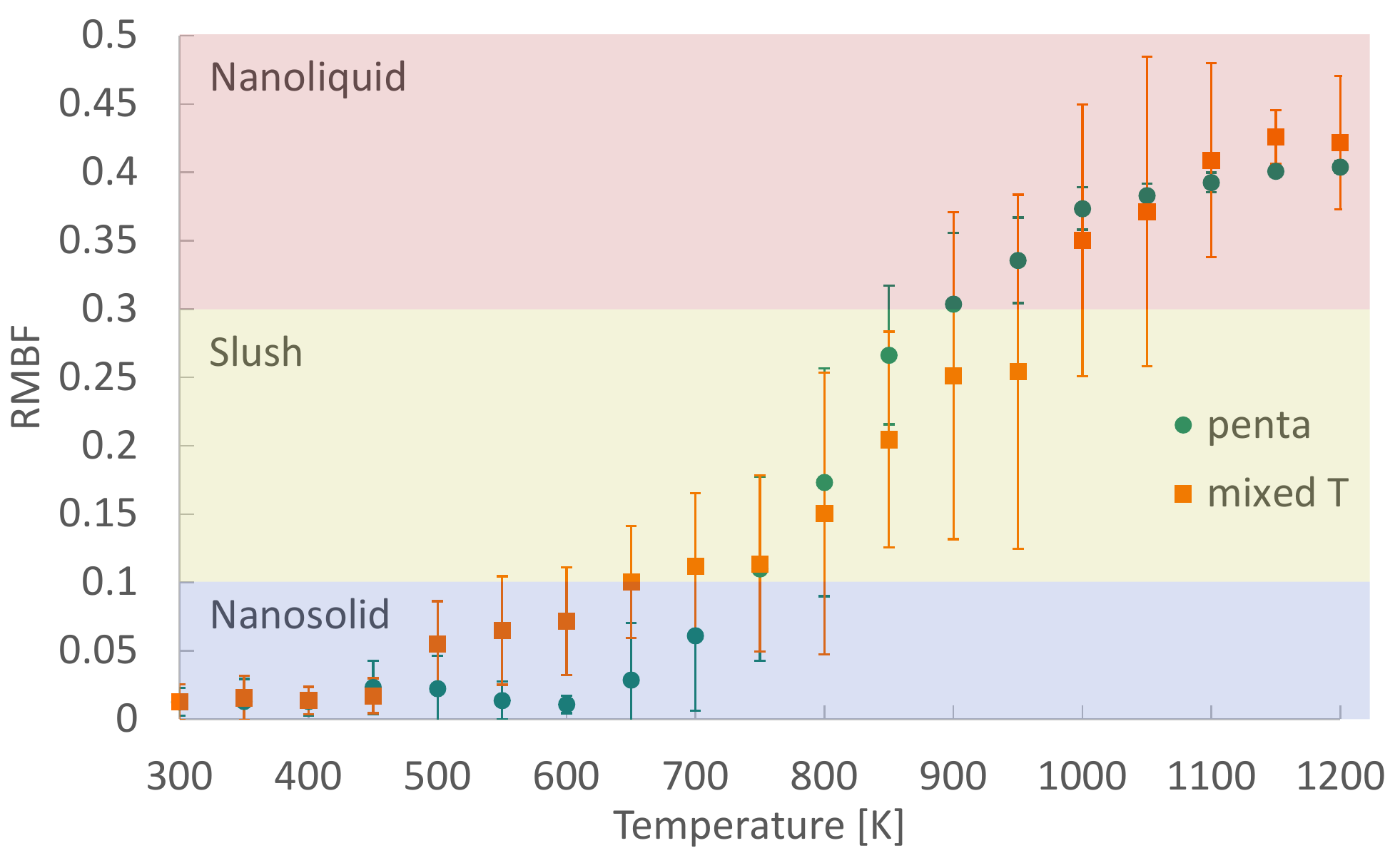}
	\caption{RMBF value against the nominal simulation temperature  for two training datasets; the error bars show the standard deviation over 20 repetitions. 
		Blue, yellow, and red shadowing highlights the characteristic values for nanosolids, slushes, and nanoliquids.}
	\label{fig:RMBF}
\end{figure}

To perform computationally inexpensive MD simulations with near-\emph{ab initio} accuracy we map the ML-FF corresponding to the two best performing 3-body kernel s (``penta" and ``mixed T onto two non-parametric M-FFs, following the procedure described in Section III C. 

We first assess their accuracy on configurations extracted from
600~K and 900~K BOMD simulations started from 3HCP and DIH initial structures, respectively.
Both systems undergo several structural changes along their trajectory. 
The computed MAEFs  for the "penta" M-FF are  0.26 $\pm$ 0.24 eV/$\text{\AA}$ for 3HCP at 600~K and  0.25 $\pm$ 0.17 eV/$\text{\AA}$ for DIH at 900~K, indicating that the M-FF retains an acceptable accuracy level while visiting configurations not represented in the ``penta'' training database. 
(Also note that higher errors should be expected for high-temperature samples, where forces have a larger modulus).

The MAEFs associated with the ``mixed T" M-FF is 0.25 $\pm$ 0.46 eV/$\text{\AA}$ for 3HCP at 600~K, and 0.17 $\pm$ 0.09 eV/$\text{\AA}$ for DIH at 900~K.

 The M-FFs described so far are appropriate for simulating dynamical runs as they contain data gathered from BOMD DFT simulations only. 
Testing their accuracy on minimized 0~K structures reveals a 0.10 $\pm$ 0.02 eV/$\text{\AA}$ MAEF for the ``penta" training set and a 0.06 $\pm$ 0.02 eV/$\text{\AA}$ MAEF for the ``mixed T" training set.
The inclusion of configurations collected during structural relaxation in the training set reduces the MAEF to 0.04 $\pm$ 0.02 eV/$\text{\AA}$, while using a many-body kernel with 4000 training points yields a further reduced 0.02 $\pm$ 0.01 eV/$\text{\AA}$ MAEF.

To further test the accuracy of our 3-body  ``penta" M-FF in a dynamical setting, we run three 200~ps-long 300~K MD simulations (see supplementary material) for each of the five geometries, and compare the PDFs and BADFs with the reference ones extracted from equally long BOMD simulations of the same structures kept at the same temperatures. 
The excellent overlap obtained for both PDFs and BADFs, visible in Figure \ref{fig:pdf}, provides some further validation of the ability of the M-FF to track the reference DFT forces.  
 The errors averaged over our five 300~K structures incurred by the M-FFs while predicting energies are 16 $\pm$ 10 meV/atom  and 9 $\pm$ 7 meV/atom  for the ``penta" and the more comprehensive``mixed T" training sets, respectively. 
A scatter plot of the energy error incurred at 900~K by the  M-FF trained on the ``mixed T" database is provided in the Supplementary Material section.

 	\subsection{Assessing the thermal behaviour of Ni$_{19}$}
To probe whether results potentially yielding novel physical insights into the nanocluster behaviour can be obtained by using a M-FF,  we finally investigate the kinetic behaviour of Ni$_{19}$, to explore the extent of shape fluctuations occurring in the nanocluster\cite{rossi2017effect, pavan2015metallic, rossi2018effect, gould2016controlling, rossi2017melting}.
To do so, we carry out MD runs at 50~K-spaced temperatures comprised between 300~K and 1200~K.
We perform four 480~ps-long simulations for each temperature for each of the five morphologies considered in this work (380 simulations in total).  
The computational speed-up factor associated with carrying out a M-FF rather than a BOMD simulation in these systems is $\sim10^{5}$. 
The root mean bond fluctuation (RMBF) is a quantity describing the average bond length oscillation at a given temperature, often used to characterize phase changes in nanoscale systems \cite{steenbergen2012electronic, ojha2013single, steenbergen2013first, steenbergen2015quantum, ojha201520, steenbergen2015two}, defined as: 
\begin{equation}
	\text{RMBF} = \frac{2}{M(M-1)} 
	\sum_{i<j} \frac{
		\sqrt{ \langle r_{ij}^{2} \rangle  - \langle r_{ij} \rangle^{2}}
	                }{ 
	                \langle r_{ij} \rangle}
	\label{eq:RMBF}, 
\end{equation}
where $M$ is the number of atoms and the averages are taken over the simulation   (excluding the first 5~ps to allow for thermal equilibration) for each atom pair.
Figure \ref{fig:RMBF} shows the RMBF value averaged over 20 simulations for each temperature value (four repeated simulations for each of the five structures) as a function of temperature for both the ``penta" and the ``mixed T" M-FFs.

The first M-FF was trained on the low temperature ``penta" database, and was thus expected to be operating in a largely extrapolatory regime. 
This force field predicted a ``slush" transition region rather than abrupt melting (cf. Fig 9, green symbols). 
The second M-FF was trained on the ``mixed T" database including configurations at 600~K and 900~K not available in the previus ``penta" database and more directly relevant to the morphologies visited by the system along the temperature ramp. 
MD simulations using this M-FF also predicted a ``slush" transition region (cf. Fig 9, orange symbols), consistent with the earlier result.

In more detail, for temperatures below 700~K, all clusters remain solid for both M-FFs, as indicated by the small ($<$ 0.1) RMBF.
In this region, the ``mixed T" M-FF alone displays a non-zero RMBF, hinting at small geometrical changes for some of the starting morphologies.

A RMBF $ > 0.3$, characteristic of nanoliquids\cite{li2008nanosolids}, is observed above 900~K in simulations using the ``penta" M-FF,  and similarly above $\sim$975~K in simulations using the``mixed T" M-FF.  
In the intermediate, approximately 700-900~K range, our Ni$_{19}$ system is predicted to be associated with a RMBF intermediate between the nanosolid and nanoliquid regimes. 
The agreement between the RMBFs of the two M-FFs in figure \ref{fig:RMBF} implies that training on low temperature structures is sufficient to predict this qualitative feature of the dynamical behaviour of the present system. 
The prediction is also consistent with the high probability of geometrical rearrangements, corresponding to slush structures, that have been discussed in details for Al systems\cite{li2008nanosolids}. 

Several detailed geometrical interconversion processes are observed during our M-FF simulations, whose in-depth characterization is ongoing and will be provided in a future work.

\vspace{1.0cm}

\section{Conclusions}
We investigated the accuracy of a Gaussian Process-based machine learning approach to the prediction of interatomic forces in metallic nanoclusters. 
In particular, we assessed the ability of different $n$-body kernels to correctly model the interactions between atoms in the Ni$_{19}$ system, and probed how the prediction accuracy is affected by ML training carried out on single-structure and multi-structure (heterogeneous) databases. 
We find that, at variance with the case of bulk Ni, a 2-body kernel is not sufficiently accurate, while a 3-body kernel is able to accurately reproduce the reference DFT forces. 
Restricting the training databases to configurations derived from a single structure yields excellent interpolation accuracy, so that a 3-body kernel can be safely used in a ``\textit{self-training}" regime.  
However, we find that ``\textit{cross-training}" the kernel is not equally successful. 
Using training databases comprising configurations derived from different cluster structures is therefore necessary to enable extrapolation, whenever this is deemed to occur e.g., by evaluation of a Kullback$-$Leibler asymmetric indicator. 

Our results suggest that mixing configurations from as few as two different structures is sufficient to train a 3-body kernel capable of robust extrapolation and thus accurate force prediction for all the structures considered.   
This in turn suggests that a 3-body kernel can achieve a very good compromise between representation power (which increases with the $n$-body kernel order), 
and speed of convergence with respect to database size (which instead decreases with $n$), provided that heterogeneous training databases are used.  
This result could be particularly useful for practical applications, since the force field predicted by a 3-body GP kernel can be mapped onto an exactly equivalent non parametric ``M-FF" force field \cite{Glielmo2018}.  
Such mapping yields a very significant efficiency gain, for all practical purposes aligning the M-FF speed of execution with that of any equivalent (e.g., 3-body) parametrised classical force field, while retaining  the accuracy and ease of training of the underlying machine learning scheme. 
As a simple feasibility test, we investigated the thermal behaviour of Ni$_{19}$ between 300~K and 1200~K. 
 To address the cluster's thermal behaviour, we carried out MD simulations, using a M-FF trained on 300~K structures and a second M-FF trained on 300, 600, and 900~K structures, adding up to a $\sim$400~ns total simulated time. 
Both M-FFs predict the occurrence of three distinct physical regimes, with similar estimates for the temperature boundaries separating them so that in particular dynamical states of the cluster, intermediate between the solid and liquid phases, are predicted to occur between 700~K and 900~K.  

\section*{Supplementary Material}
See supplementary material for details on the BOMD and M-FF simulations, the learning curve graphs for DIH, BIP, 4HCP and dDIH cluster structures, more informations about the Kullback$-$Liebler divergence method, the complete set of graphs for the MAEFS incurred by the 3- and many-body kernels while \textit{mixed-training}, a description of the Monte Carlo Metropolis simulations performed, and the scatter plot of energy predictions for the ``mixed T" M-FF on 1500 configurations extracted from DFT BOMD simulations at 900~K.
\section*{Acknowledgements}
CZ and AG acknowledge funding by the Engineering and Physical Sciences
Research Council (EPSRC) through the Centre for Doctoral Training Cross Disciplinary Approaches to Non-Equilibrium Systems (CANES, Grant No. EP/L015854/1) and by the Office of Naval Research Global (ONRG Award No. N62909-15-1-N079). 
ADV acknowledges further support by the EPSRC HEmS Grant No. EP/L014742/1 and, together with AF, by the European Union\textquoteright s Horizon 2020 research and innovation program (Grant No. 676580, The NOMAD Laboratory, a European Centre of Excellence). 
We are grateful to the UK Materials and Molecular Modelling Hub for computational resources, which is partially funded by EPSRC (EP/P020194/1).
FB acknowledges financial support from the UK Engineering
and Physical Sciences Research Council (EPSRC),
under Grants No. EP/GO03146/1 and No. EP/J010812/1.
KR acknowledges financial support from EPSRC, Grant No.
ER/M506357/1, the Thomas Young Centre, the MacDiarmid Institute, 
and Auckland University. 
KR and FB acknowledge the financial support offered by the Royal Society
under the project number RG120207.
KR also thanks Krista Grace Steenbergen for useful discussions.
The authors wish to acknowledge the contribution of NeSI high-performance computing facilities to the
results of this research. 
NZ's national facilities are provided by the NZ eScience Infrastructure and
funded jointly by NeSI's collaborator institutions and through the Ministry of Business, Innovation and
Employment's Research Infrastructure programme. URL https://www.nesi.org.nz.

\bibliography{bib_tot}

\break
\newpage

\section*{Supplementary Material}
\subsection*{DFT simulation details}
All our Density Functional Theory (DFT) simulations are performed using the VASP package \cite{kresse2001vienna}. 
The plane-wave energy cut-off is set to 270~eV, while the energy convergence cut-off is $10^{-4}$~eV. 
The exchange-correlation functional is calculated using the PBE/GGA approximation \cite{perdew1996generalized} and spin-orbit coupling effects are accounted for. 
Each calculation is carried out using periodic boundary conditions in $6a \times 6a \times 6a$ cubic simulation cell, where $a$ is the Ni bulk lattice parameter (3.52 $\text{\AA}$), which is sufficient to avoid spurious interactions between images of the system. 
A Nose-Hoover thermostat is used to control the temperature during NVT BOMD runs at 300~K; data at 600~K and 900~K are also gathered for further comparison. 
The Newton's equations of motion are integrated with a timestep of 3~fs.
\subsection*{Learning curves}
We report in Figure \ref{fig:learning_curves} the learning curves for the DIH, BIP, 4HCP, and dDIH cluster morphologies while trained on a 2-, 3-, and a many-body kernel.
\begin{figure*}
\includegraphics[trim = 0.8cm 0 0 0.5cm , clip, width=0.48\linewidth]{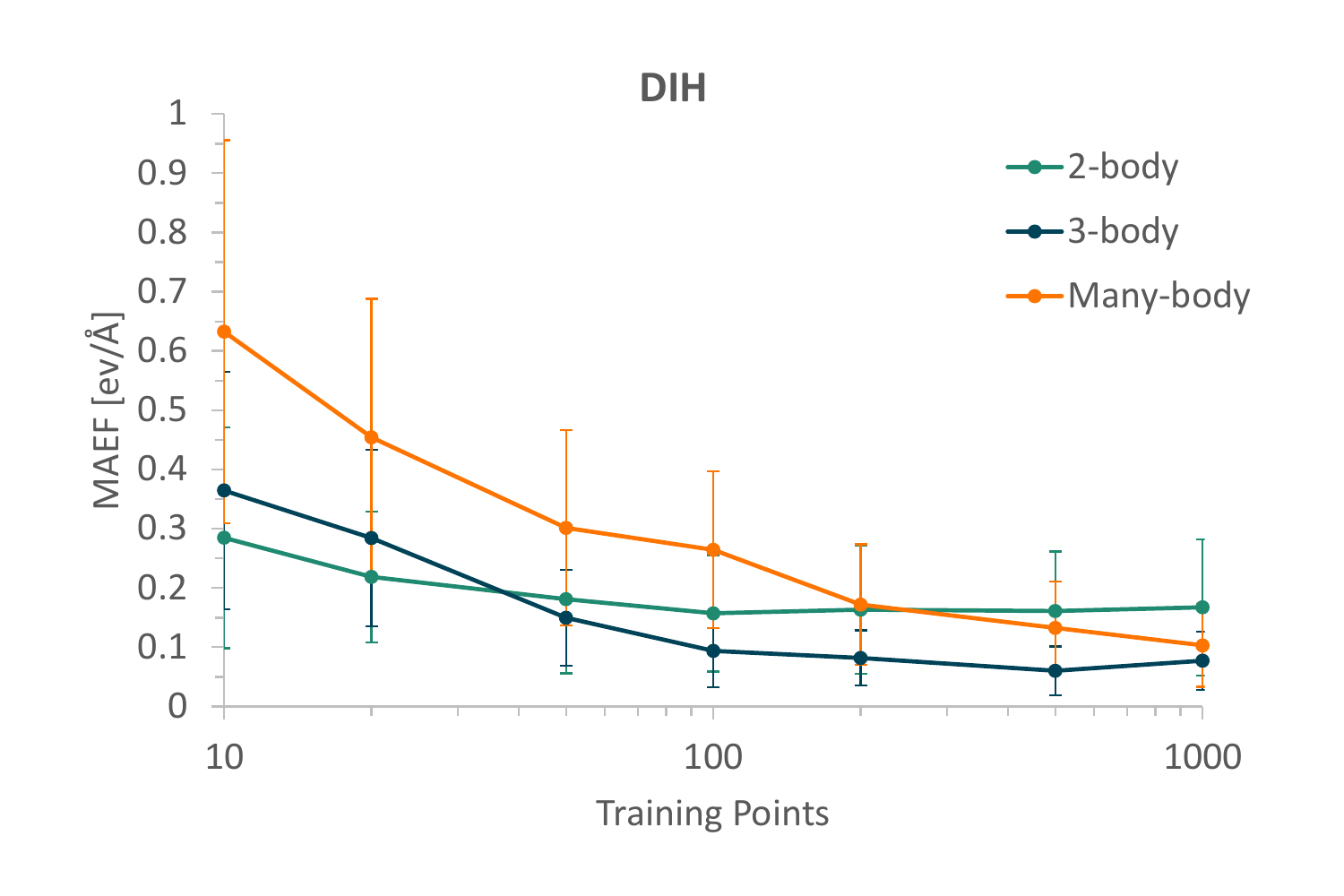}
\includegraphics[trim = 0.8cm 0 0 0.5cm , clip,width=.48\linewidth]{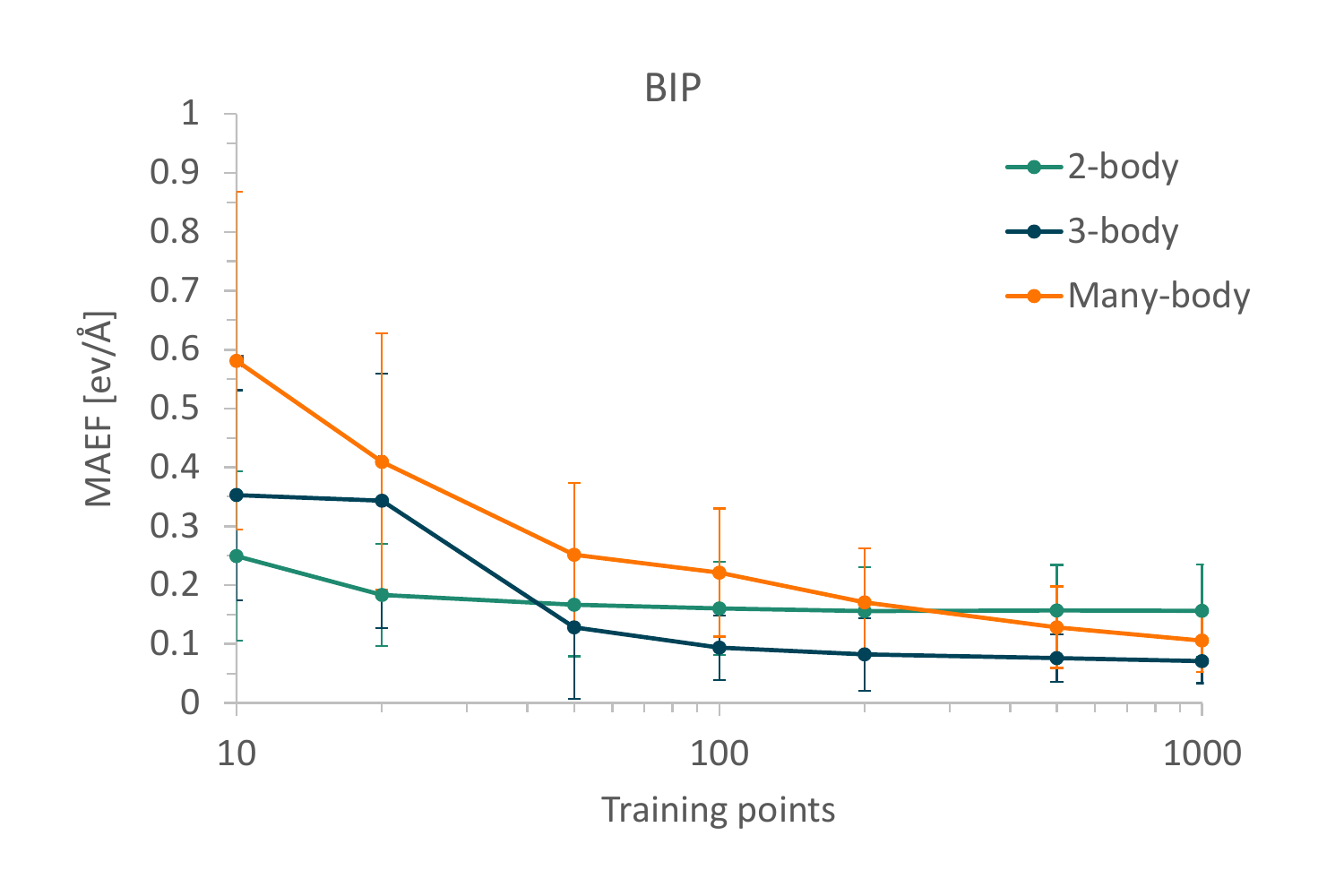}
\includegraphics[trim = 0.8cm 0 0 0.5cm , clip,width=.48\linewidth]{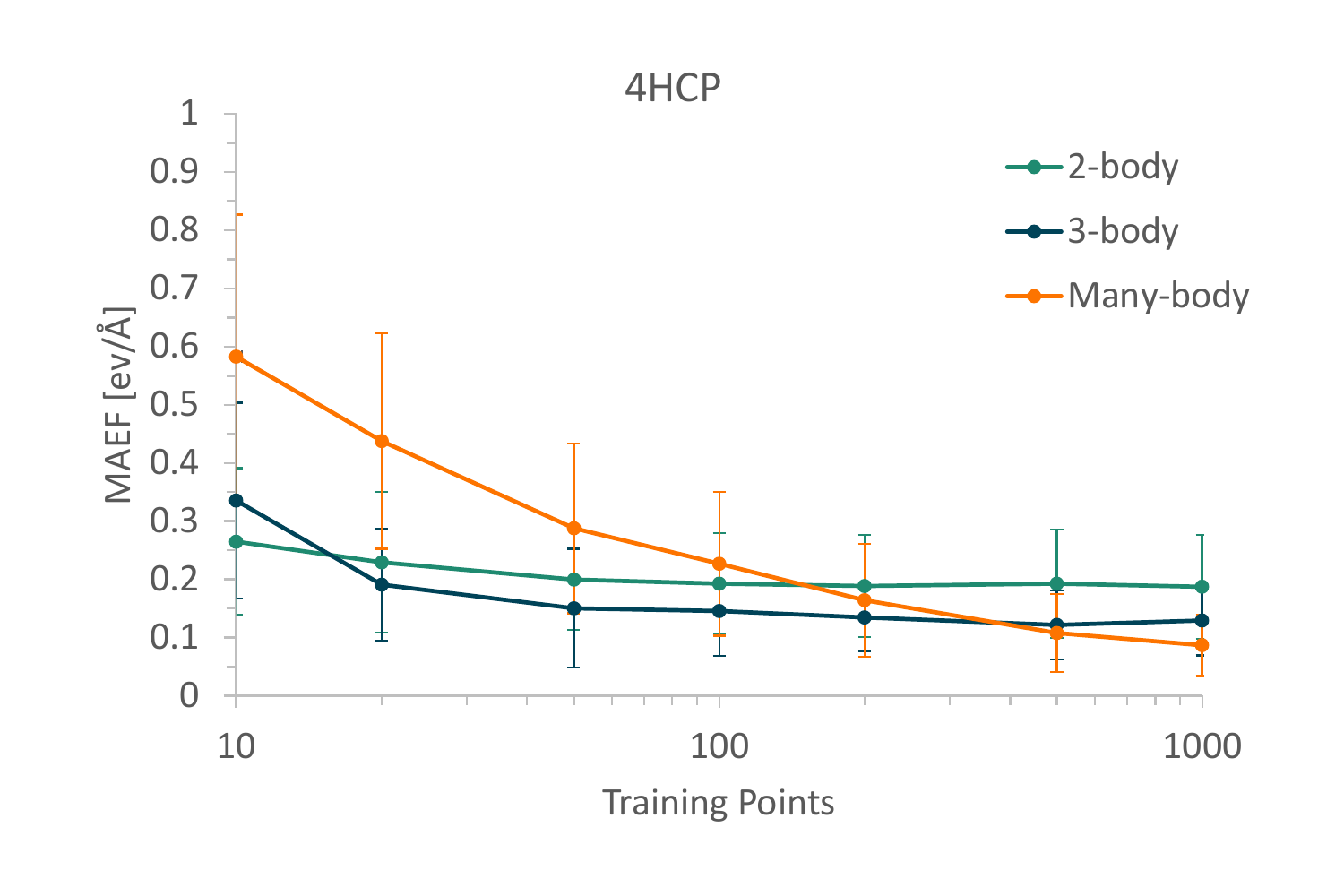}
\includegraphics[trim = 0.8cm 0 0 0.5cm , clip,width=.48\linewidth]{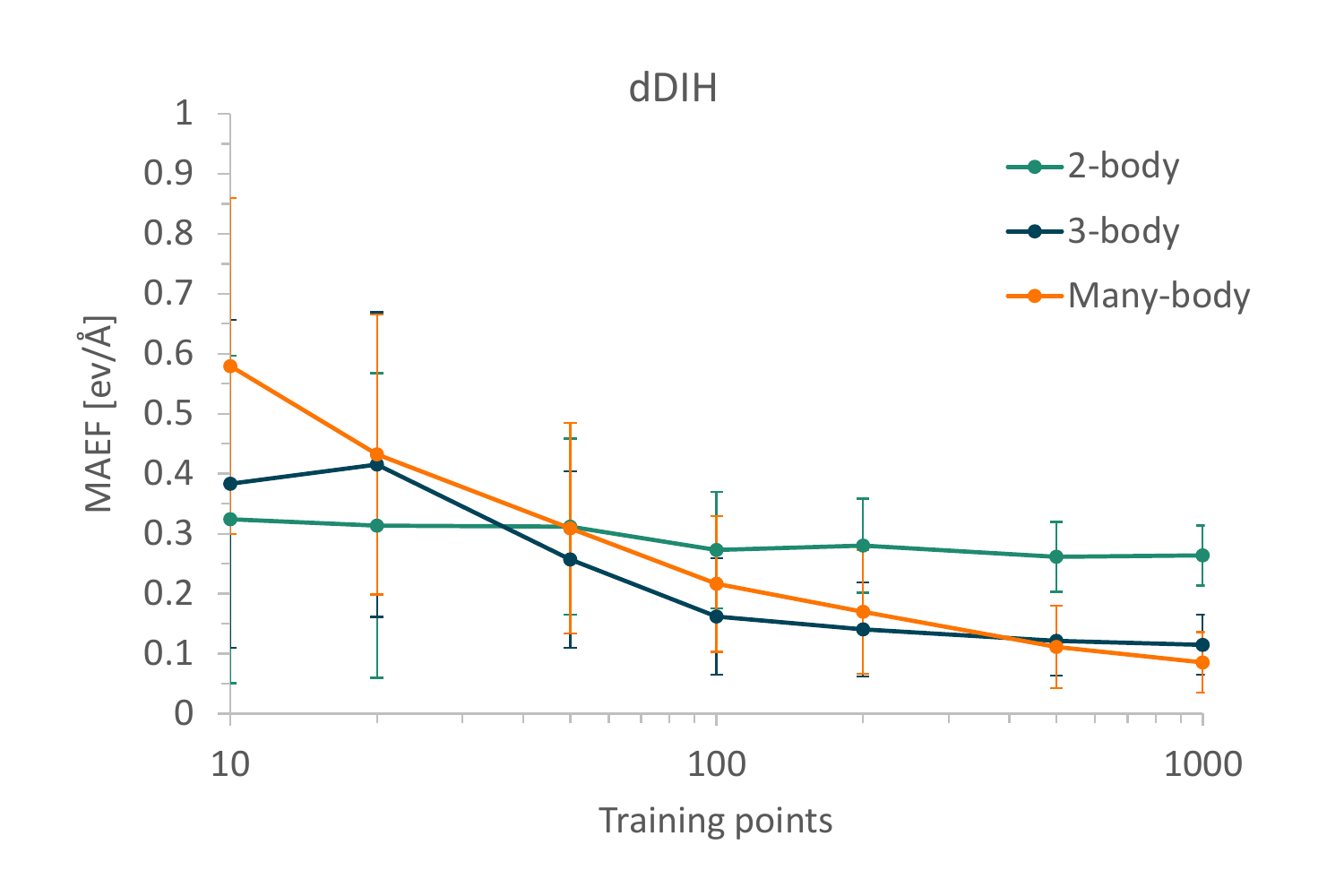}
	\caption{Learning curves for training and testing on a DIH, BIP, HCP and a dDIH cluster structures. 
		The kernel's $n$-body order is colour coded, with 2-body in cyan, 3-body in blue, and many-body in orange.}
	\label{fig:learning_curves}
\end{figure*}
\subsection*{KL divergence and cross-errors normalization}

\begin{figure}[b]
\includegraphics[width=1.0\linewidth]{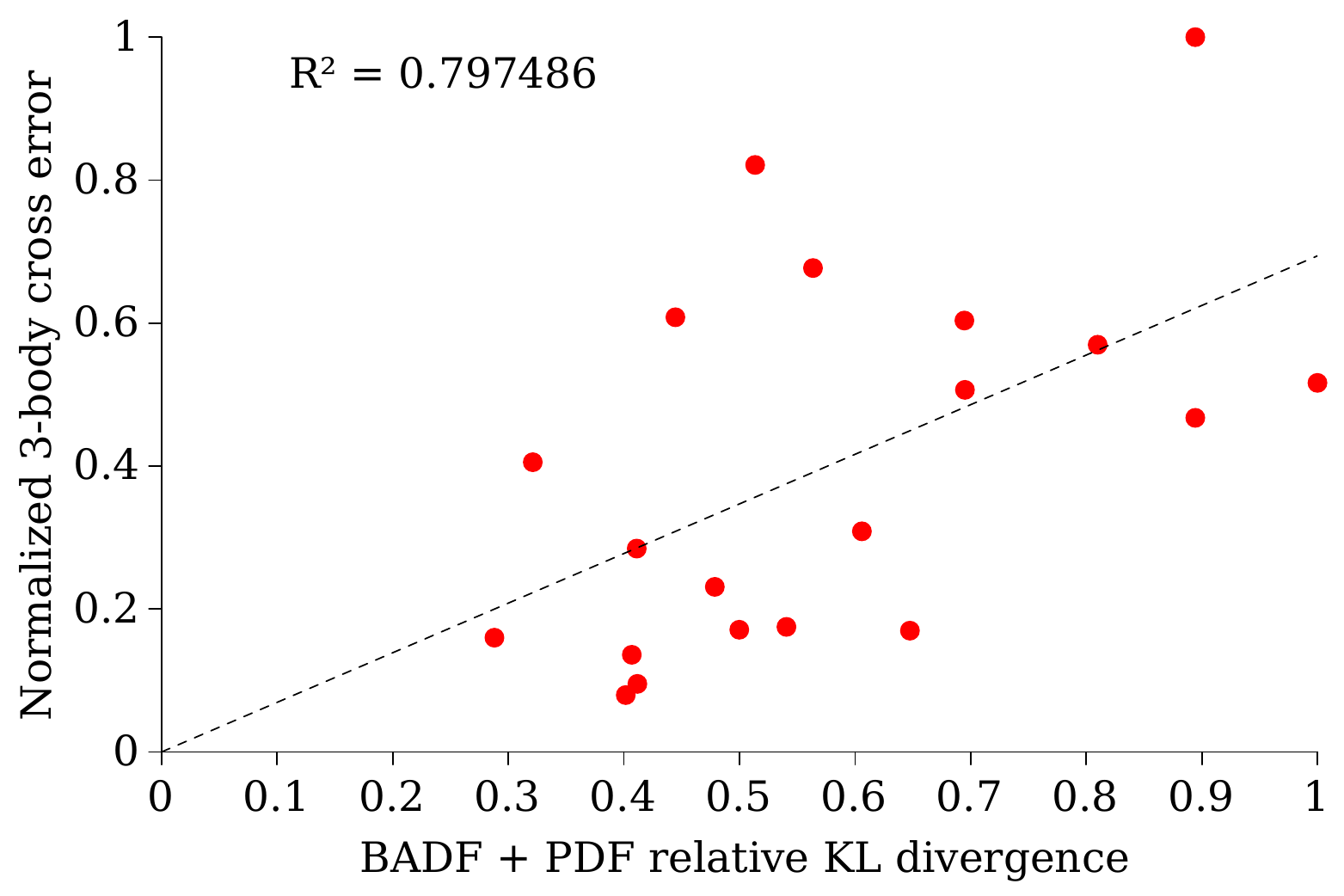}
\caption{Scatter plot highlighting the correlation between
the relative KL divergence of the five configuration weighed functions 2/3 BADFs + 1/3 PDFs and the 3-body cross-learning normalized MAEFs (see text).}
\label{fig:3BKLD}
\end{figure}

In order to check for a correlation between the cross-testing 2- and 3-body MAEFs and the KL divergence calculated for PDFs and BADFs of the five configurations, we first normalize both sets of results to values comprised between 0 and 1. 
In the case of the 2- and 3-body kernels, this implies subtracting the self-training MAEFs from each set in order to obtain a zero for every self-training MAEF, before dividing all values by the maximum value. 
For the KL divergence of both BADFs and PDFs, we divide by the maximum value obtained in order to, again, rescale to the [0:1] range.
Figure \ref{fig:3BKLD} contains the scatter plot between the 3-body normalized cross-error and the corresponding KL divergence values associated with the twenty possible ordered structure couples obtained starting from our five cluster structures. 
Here the function fed to the KL estimator is, for each cluster structure, a (still normalized) weighed sum of the structure's PDF (weight = 1/3) and BADF (weight = 2/3).  

\subsection*{Mixed datasets MAEFs}
\begin{figure*}
\centering
\includegraphics[width = 17cm, height = 11cm]{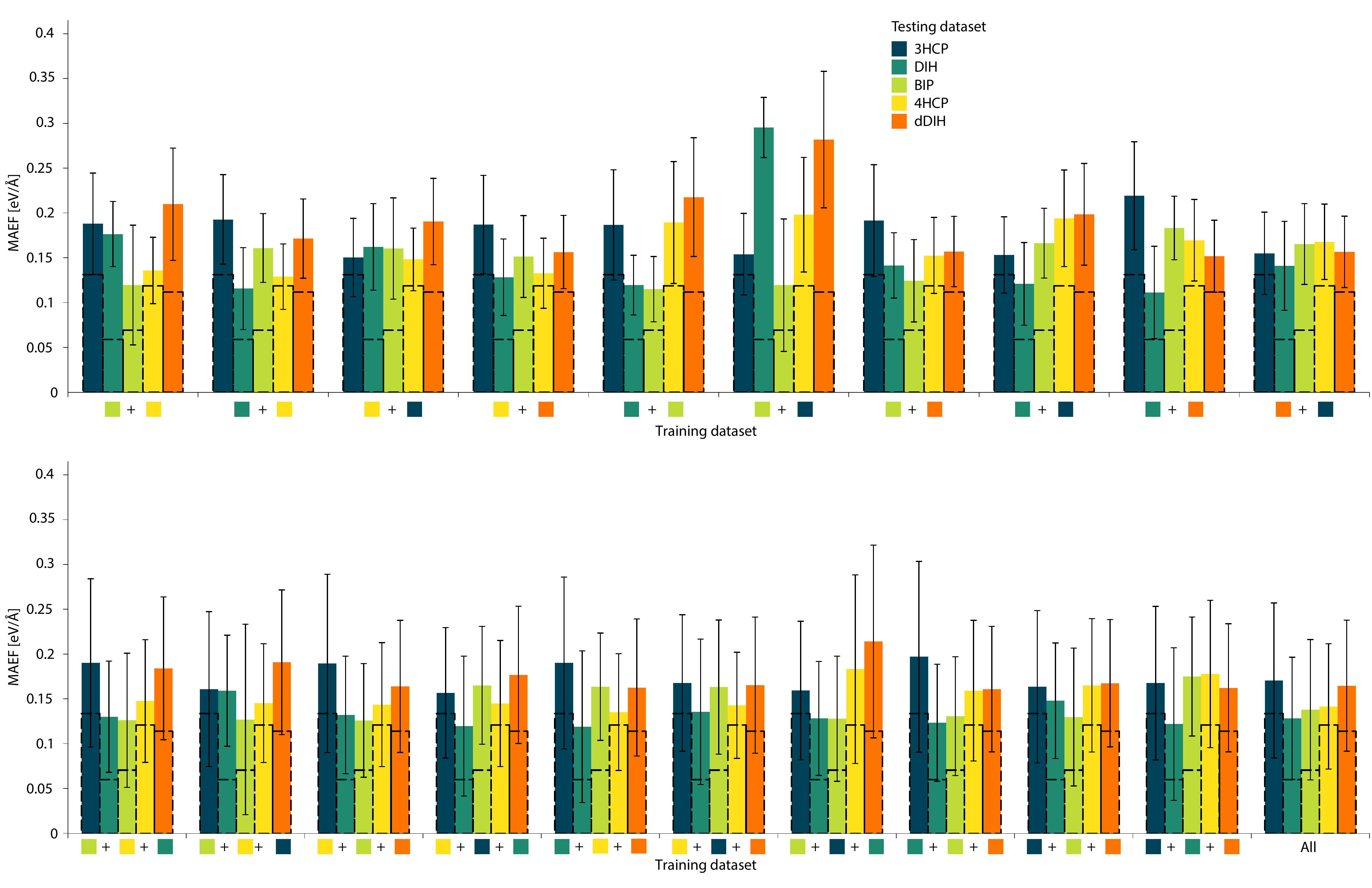}
\caption{MAEF and its standard deviation for the 3-body kernel trained on mixed databases compound of two (top) and three or all (bottom) structures with $N_{TR}$ = 500 and tested on all our five Ni structures at 300~K. Self-training accuracies are reported in black dotted lines for reference.}
\label{fig:three_shards_err}
\end{figure*}

\begin{figure*}
\centering
\includegraphics[width = 17cm, height = 11cm]{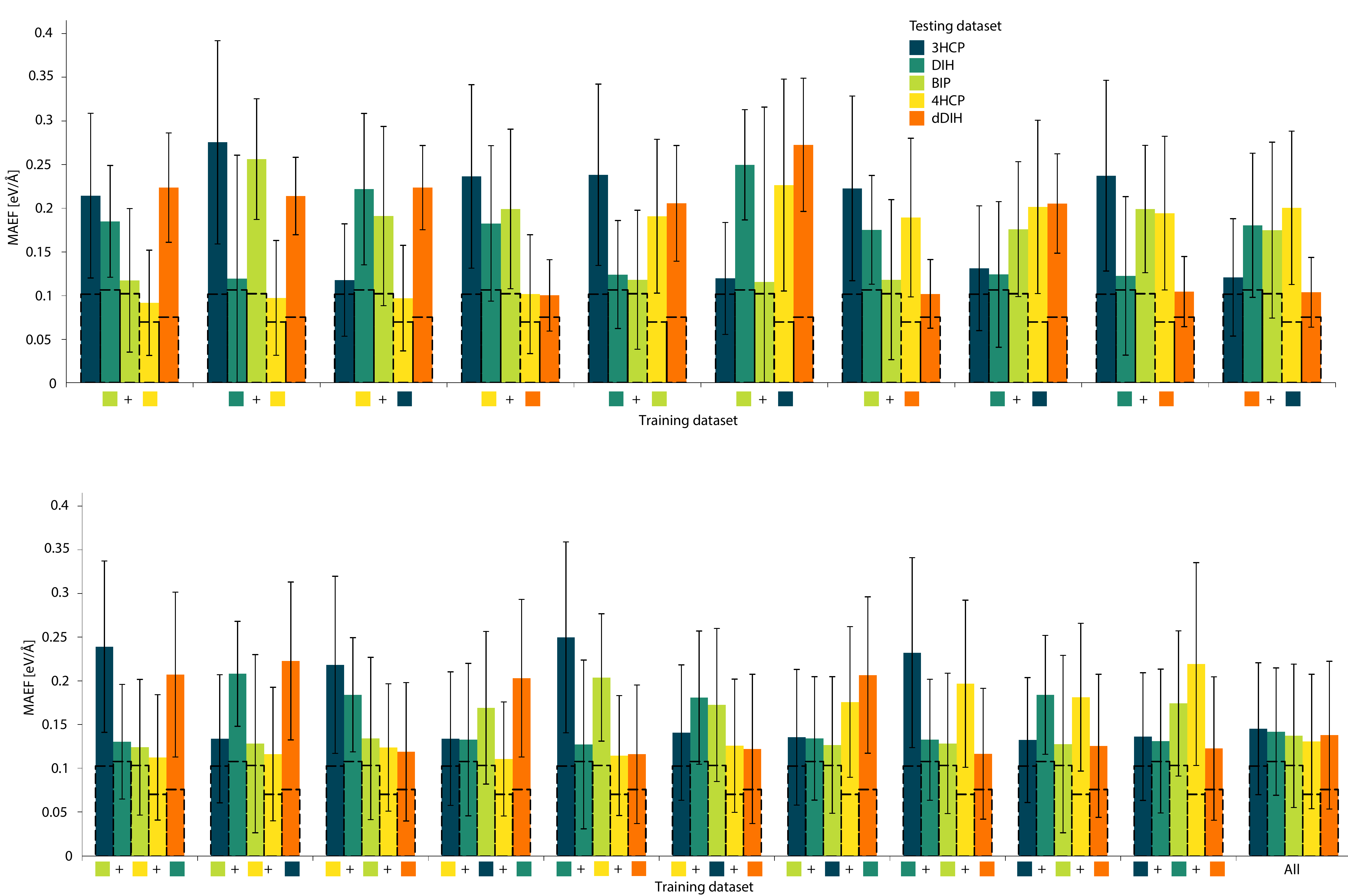}

\caption{MAEF and its standard deviation for the many body kernel trained on mixed databases compound of two (top) and three or all (bottom) structures with $N_{tr}$ = 1000  and tested on the five Ni structures at 300~K. Self-training accuracies are reported as black dotted lines for reference.}
\label{fig:many_shards_err}
\end{figure*}

Figures \ref{fig:three_shards_err} and \ref{fig:many_shards_err} illustrate the errors incurred by the 3- and many-body kernels when trained on every possible combination of two, three and five configurations.
As a highlight, Figure \ref{fig:three_shards_err} reports an example of the performance of the dDIH+3HCP training database. 
This is the best-performing two-structure training set for the 3-body kernel, although this heterogeneous training set does not contain the 4HCP structure, which is the ``best'' structure for cross-testing.  
This can be rationalized by analyzing the PDFs and BADFs of the two configurations (see Figure 5 of the main article), noting that bond distance and angle values not contained in the PDF and BADF of one structure are present in the other, and viceversa.

\subsection*{Monte Carlo algorithm for database selection}
To estimate the error reduction achievable by optimising the data point selection in the $\mathcal{D}_{tr}$, we compare randomly built training datasets with the best performing sets we find using a Monte Carlo algorithm for the same $N_{tr}$.

In order to select the training database which yields the minimum average prediction error while also keeping the accuracy relatively homogeneous, we chose to minimise the following linear combination of the MAEF and the standard deviation of the absolute error, both taken as functions of the training set: 
\begin{equation}
X(\mathcal{D}_{tr}) = \text{MAEF}(\mathcal{D}_{tr}) + 2  \sigma({\mathcal{D}_{tr}}).
\end{equation}
The Monte Carlo optimization was structured as follows. 
First, we built and kept fixed a testing set $\mathcal{D}_{test}$ which we used to test the performance of all GPs.
We then initialialized a training database of $N_{tr}$ randomly chosen configurations and carried out Metropolis steps swapping database configurations with new ones, randomly chosen from our complete configuration pool.   
This optimization process typically yields a MAEF lower by just 0.001 eV/$\text{\AA}$ than the MAEF associated to a randomly chosen training set when using $N_{tr}=200$ data points to train the many-body kernel, a result further insensitive to increasing $N_{tr}$.
\subsection*{M-FF simulation details}
The M-FF simulations are carried out using the Atomic Simulation Environment (ASE) python package \cite{larsen2017atomic} and a Langevin thermostat to control the temperature (with gamma parameter set at 0.001), using a 3~fs time step.  
\subsection*{Energy prediction accuracy}
\begin{figure}[h]
\includegraphics[width=1.0\linewidth]{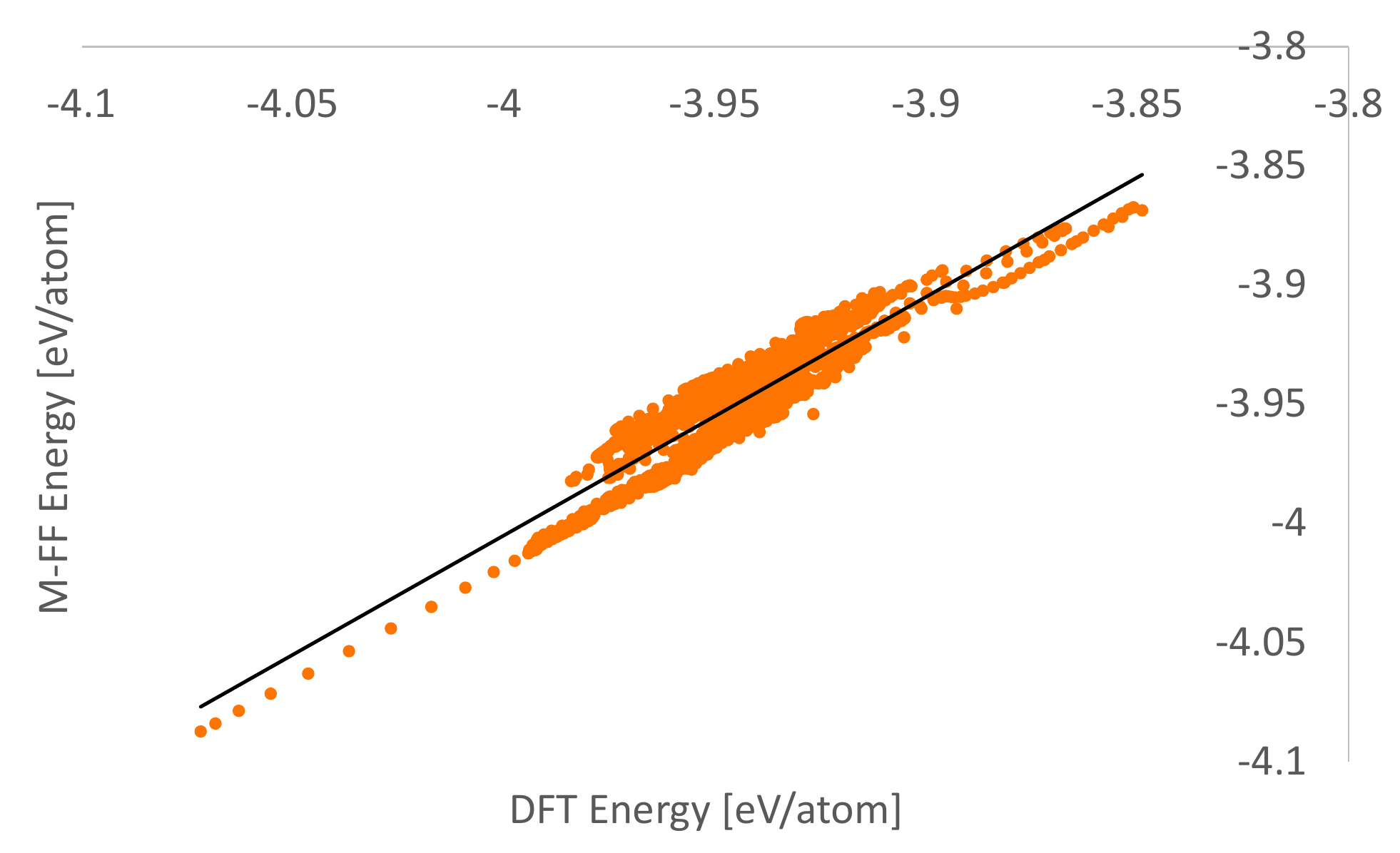}
\caption{Scatter plot displaying the energy error incurred by the  M-FF trained on the ``mixed T" described in the main text. The test is carried out on 1500 configurations extracted from a DFT BOMD simulation at 900~K. The black line provided as a guide for the eye indicates a perfect fit.}
\label{fig:energy_scatter}
\end{figure}

Finally, we report in Figure \ref{fig:energy_scatter} a scatter plot for the energy predictions obtained from the ``mixed T" M-FF on a set of 1500 testing configurations extracted from a DFT BOMD simulation carried out at 900~K (see also main manuscript text). The mean energy error is 13 $\pm$ 13 eV/atom.

\end{document}